\documentclass[11pt]{article}
\usepackage[margin=0.8in]{geometry}
\usepackage{amsmath,amssymb,bm}
\usepackage{siunitx}
\usepackage{booktabs}
\usepackage{hyperref}
\usepackage{caption}

\usepackage{graphicx}
\newcommand{\Tr}{\operatorname{Tr}}

\title{Purcell-Like Environmental Enhancement of Classical Antennas:\\ Self and Transfer Effects}
\author{Alex Krasnok\\
Department of Electrical and Computer Engineering, Florida International University\\
\texttt{akrasnok@fiu.edu}}
\date{}

\begin{document}
\maketitle

\begin{abstract}
Environmental “range boosts” in wireless links are often explained through radiation-pattern intuition, yet the underlying physics is more cleanly captured by two environment-controlled quantities: \emph{radiative damping} of the radiator and \emph{channel coupling} between transmitter and receiver.  Building from a dyadic-Green-function current--field formulation, we introduce an operational two-factor description of Purcell-like behavior for classical antennas.  A \emph{self} factor quantifies environment-induced changes in radiative damping under an explicit excitation convention (e.g., $F_{\mathrm{self}}^{(I)}=R_{\mathrm{rad}}/R_{\mathrm{rad},0}$), while a \emph{transfer} factor quantifies environment-induced changes in Tx--Rx coupling (e.g., $F_{\mathrm{tr}}=|h|^{2}/|h_{0}|^{2}$, equivalently expressible via cross-Green functions or mutual impedance).  We provide measurement-aware extraction workflows (VNA $S_{11}\!\rightarrow Z_{\mathrm{in}}$ with efficiency and realized-gain accounting; link-test normalization to isolate $F_{\mathrm{tr}}$) and falsification diagnostics that prevent conflating true radiative enhancement with mismatch or added absorption.  Finally, we translate self/transfer modifications into link-budget and range scalings and illustrate the framework across practical environments from VHF to mmWave, including platforms/ground planes, body proximity, field-expedient environmental radiators, terrain and passive redirection, tunnel/canyon confinement, and engineered scattering environments such as reflectarrays, metasurfaces, and reconfigurable intelligent surfaces (RIS).
\end{abstract}

\section{Introduction}
The \emph{Purcell effect}---the modification of an emitter’s radiative decay by its electromagnetic environment---is, at its core, a statement about \emph{radiative damping}: the surrounding electromagnetic landscape reshapes the dissipative radiative channels available to a source and thereby changes the rate at which the source can shed energy as radiation \cite{purcell1946,kleppner1981,haroche1988}.  In cavity QED and nanophotonics, this principle is routinely exploited by engineering modal structure, confinement, and leakage (radiative and absorptive) \cite{kleppner1981,yablonovitch1987,joulain2003,novotny2012nano}.  A historically clear illustration comes from Drexhage-type interface experiments, where the decay rate oscillates with distance to a boundary due to interference between direct and reflected (and surface-bound) fields \cite{drexhage1970,chpinoptics1978}.

A particularly powerful and geometry-agnostic language for these effects is the dyadic Green tensor $\bm{G}$ (and its density-of-states interpretations), which encodes how materials and boundaries redistribute both radiative and nonradiative dissipation channels \cite{sipe1987,joulain2003,carminati2015ssr}.  For an electric dipole at $\bm{r}$ one often introduces an LDOS proportional to the imaginary part of the on-site Green tensor,
\begin{equation}
\rho(\bm{r},\omega)=\frac{2\omega}{\pi c^2}\,\Im\!\left[\Tr\,\bm{G}(\bm{r},\bm{r};\omega)\right],
\label{eq:ldos}
\end{equation}
emphasizing that $\Im\{\bm{G}(\bm{r},\bm{r})\}$ governs radiative damping in the weak-coupling regime.  Importantly for \emph{links}, the same operator also governs \emph{two-point} coupling: off-diagonal (cross) Green-tensor components determine how efficiently energy launched at $\bm{r}_t$ reaches $\bm{r}_r$.  In nano-optics this two-point connectivity can be formalized through cross-density-of-states (CDOS) concepts that quantify intrinsic spatial coherence in complex environments \cite{caze2013prl,carminati2015ssr}.

The same physics has an exact classical counterpart for driven antennas.  A resonant antenna is a radiating oscillator whose energy-loss rate includes radiative damping; modifying the environment modifies that damping.  In classical antenna theory this viewpoint is embedded in reaction/reciprocity formalisms \cite{rumsey1954,harrington1961,richmond1961}.  For a time-harmonic current distribution $\bm{J}(\bm{r})$, the time-averaged power delivered to the fields can be written as
\begin{equation}
P_{\mathrm{in}}=\frac{1}{2}\Re \!\int_V \bm{J}^\ast(\bm{r})\cdot \bm{E}(\bm{r})\,dV,
\label{eq:pin}
\end{equation}
with the field expressed through the dyadic Green tensor,
\begin{equation}
\bm{E}(\bm{r})=j\omega\mu \int_V \bm{G}(\bm{r},\bm{r}';\omega)\cdot \bm{J}(\bm{r}')\,dV',
\qquad
\bm{G}=\bm{G}_0+\bm{G}_{\mathrm{sc}}.
\label{eq:green}
\end{equation}
Equations \eqref{eq:pin}--\eqref{eq:green} make explicit that boundaries, nearby objects, and structured media enter \emph{only} through $\bm{G}_{\mathrm{sc}}$ and therefore modify the work done by (and on) the antenna currents.

\begin{figure}[!t]
\centering
\includegraphics[width=\linewidth]{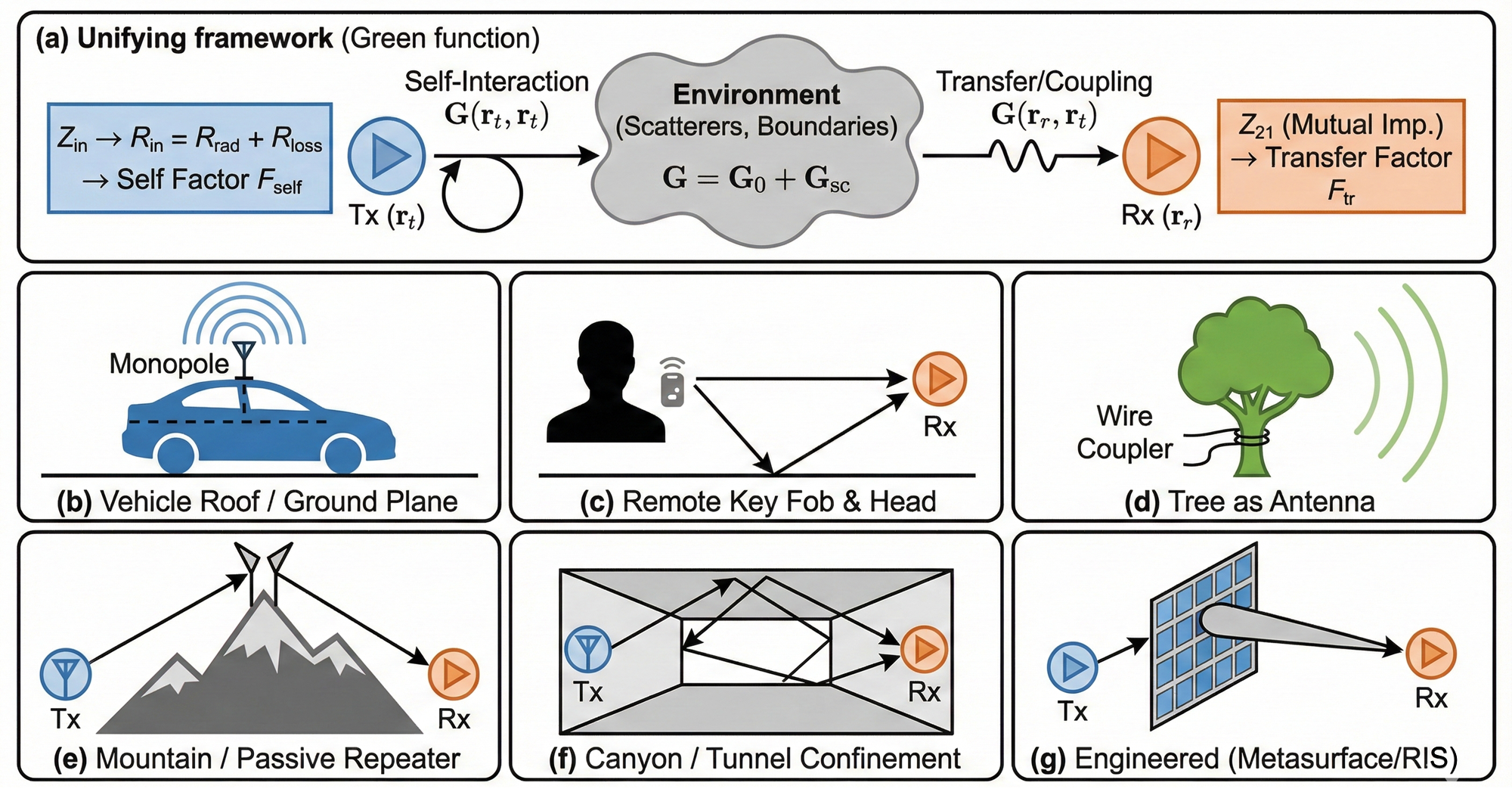}
\caption{\textbf{Environment-assisted radiation and coupling: two classical Purcell-like factors.}
(a) Unifying framework: the environment enters through $\bm{G}=\bm{G}_0+\bm{G}_{\mathrm{sc}}$.
Self-interaction via $\bm{G}(\bm{r}_t,\bm{r}_t)$ modifies $Z_{\mathrm{in}}$ and radiative damping (self factor $F_{\mathrm{self}}$),
while cross-interaction via $\bm{G}(\bm{r}_r,\bm{r}_t)$ controls Tx--Rx coupling (transfer factor $F_{\mathrm{tr}}$),
equivalently described by mutual impedance or CDOS \cite{rumsey1954,richmond1961,caze2013prl}.
(b)--(g) Representative scenarios treated in the text: platforms/ground planes, body proximity, trees as environmental radiators, mountains/passive repeaters, canyon/tunnel confinement, and engineered environments (reflectarrays/metasurfaces/RIS).}
\label{fig:fig1}
\end{figure}

This operator statement has been especially fruitful in nanoantenna science, where metallic and dielectric resonators provide compact, mode-engineered radiators and receivers.  The optical-nanoantenna literature has long emphasized that directivity, impedance matching, and radiative/nonradiative partitioning are all manifestations of the same environment-shaped channel structure \cite{krasnok2013nanoantennas}.  In all-dielectric platforms, magnetic and electric Mie resonances enable low-loss nanoantennas and strongly directive emission, and have been verified experimentally \cite{filonov2012alldielectric,krasnok2014superdirective}.  These developments also enabled explicit demonstrations of large Purcell enhancements in dielectric systems (including magnetic-dipole analogues in wire metamaterials and strong Purcell enhancement in all-dielectric chains), reinforcing that ``Purcell'' language is not confined to plasmonic hot spots \cite{slobozhanyuk2014magpurcell,krasnok2016enhpurcell}.

A complementary line of work has made the antenna--Purcell analogy \emph{operational} in the RF/microwave domain by linking Purcell-like enhancement to measurable port quantities.  In particular, the Purcell factor for dipole-like electric and magnetic antennas can be retrieved from input impedance (or radiation resistance under a stated excitation convention), and experimentally validated using standard VNA measurements \cite{krasnok2015antennaPurcell}.  These results establish the conceptual bridge: LDOS/Green-tensor control in optics and radiation-resistance control in antennas are two faces of the same environment-modified radiative damping.

\textbf{Scope and what is new in this work.}
Despite these foundations, a persistent practical ambiguity remains in \emph{wireless links}: many claimed ``environmental enhancements'' are inferred from received power, coverage, or range, where the same dB-level change can originate from two distinct mechanisms:
(i) \emph{local} changes at the transmitter/receiver (matching, efficiency, directivity) and
(ii) \emph{transfer} changes in the propagation channel (multipath interference, guiding, scattering, redirection).
The present manuscript contributes a measurement-facing framework that separates these mechanisms in a way that is directly testable:
we introduce two complementary Purcell-like factors extracted from the same Green-tensor formalism---a \emph{self} factor tied to the self-interaction $\bm{G}(\bm{r}_t,\bm{r}_t)$ and observable through $Z_{\mathrm{in}}$ (with explicit attention to $R_{\mathrm{rad}}$ versus $R_{\mathrm{loss}}$), and a \emph{transfer} factor tied to the cross-interaction $\bm{G}(\bm{r}_r,\bm{r}_t)$ and observable through mutual coupling or a normalized channel coefficient.  We then translate these factors into link-budget and range scalings, and we provide falsification-oriented reporting guidance (e.g., simultaneous $S_{11}$ monitoring to prevent mislabeling transfer fading as ``antenna gain'').

Figure~\ref{fig:fig1} provides a roadmap for the environments considered.  Panel~(a) summarizes the unifying viewpoint and the two complementary quantities: (i) \emph{self} modification of radiative damping governed by $\bm{G}(\bm{r}_t,\bm{r}_t)$ and observable through $Z_{\mathrm{in}}$ and its decomposition $R_{\mathrm{in}}=R_{\mathrm{rad}}+R_{\mathrm{loss}}$, and (ii) \emph{transfer} modification of coupling governed by $\bm{G}(\bm{r}_r,\bm{r}_t)$ and observable through mutual coupling/link gain, with a natural connection to CDOS-type two-point coherence measures \cite{richmond1961,caze2013prl}.

The remainder of the paper is organized as follows.  Section~2 formulates the current--field integral and Green-function description in a way that makes the \emph{excitation convention} explicit (fixed current versus finite source impedance) and separates radiative damping from absorption by writing $R_{\mathrm{in}}=R_{\mathrm{rad}}+R_{\mathrm{loss}}$.  In the same framework we introduce the transfer factor through the cross-Green tensor and mutual coupling, connecting it to link gain and to coherence-based measures such as CDOS.  Subsequent sections treat the scenarios in Fig.~\ref{fig:fig1}(b)--(g) as quantitative case studies with regime-appropriate scaling laws and end with practical measurement guidance for extracting genuine self enhancement and for diagnosing transfer-dominant improvements.

\section{Green-function and impedance viewpoints}

\begin{figure}[t]
\centering
\includegraphics[width=\linewidth]{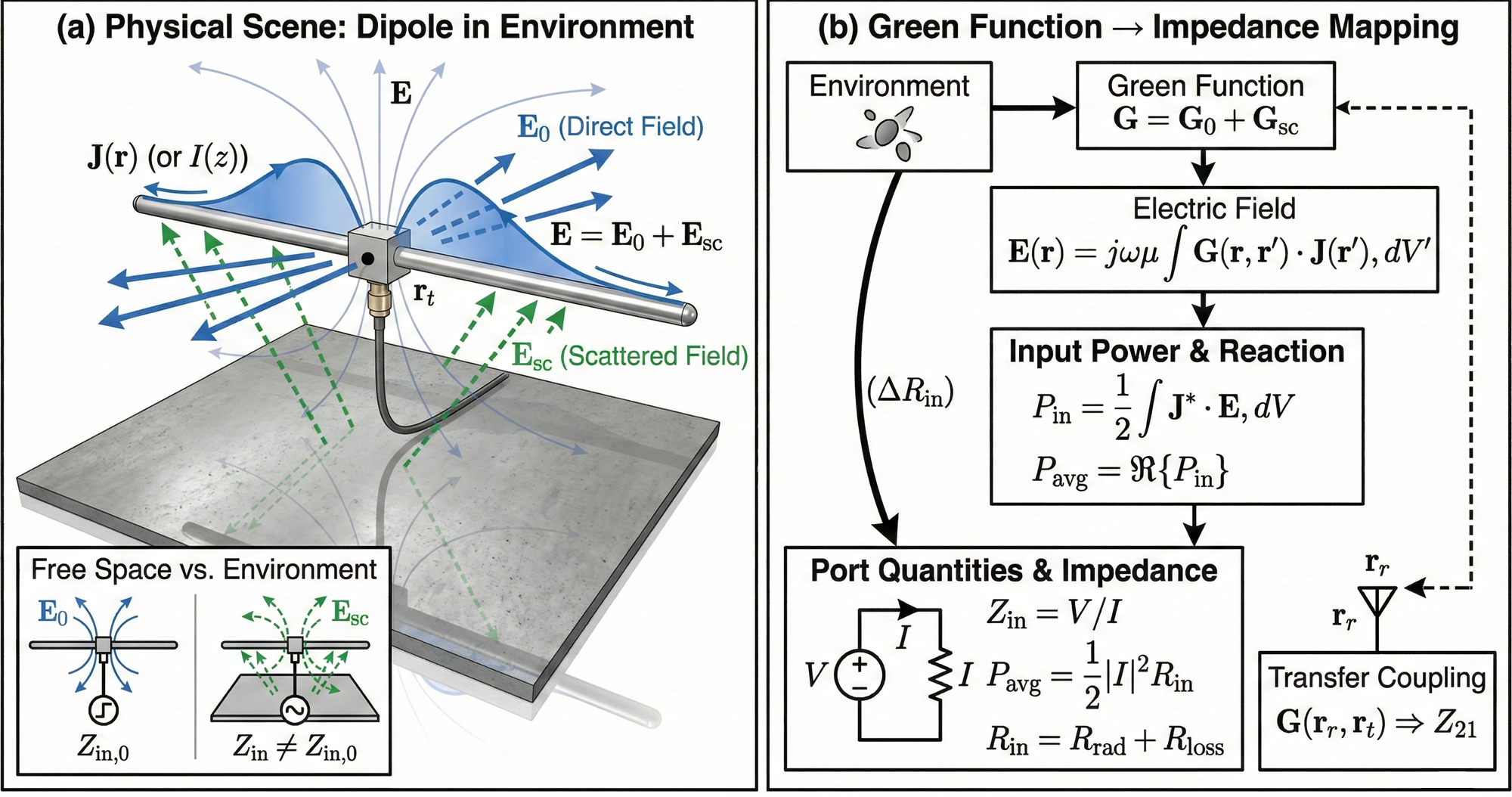}
\caption{\textbf{Green-function and impedance viewpoints for a realistic dipole in an environment.}
(a) Physical scene: a center-fed dipole with an illustrative sinusoidal current distribution $I(z)$ (or $\bm{J}(\bm{r})$) placed near a nearby structure (e.g., a conducting surface). The total field is decomposed into a direct contribution and an environmentally scattered contribution, $\bm{E}=\bm{E}_0+\bm{E}_{\mathrm{sc}}$, which modifies the port impedance relative to free space ($Z_{\mathrm{in}}\neq Z_{\mathrm{in},0}$).
(b) Mapping: the environment enters through $\bm{G}=\bm{G}_0+\bm{G}_{\mathrm{sc}}$; $\bm{G}$ maps currents to fields, fields to reaction power, and reaction power to port quantities ($Z_{\mathrm{in}}$, $R_{\mathrm{in}}=R_{\mathrm{rad}}+R_{\mathrm{loss}}$). The same operator controls transfer coupling between a transmitter at $\bm{r}_t$ and a receiver at $\bm{r}_r$ through the cross-Green function $\bm{G}(\bm{r}_r,\bm{r}_t)$ (often expressed as a mutual coupling/impedance quantity).}
\label{fig:fig2}
\end{figure}

\subsection{Time-harmonic power, reaction, and environment-induced impedance}
Figure~\ref{fig:fig2} summarizes the logic chain used throughout this review. Panel~(a) shows a realistic center-fed dipole near an environment (here a conducting surface), with an illustrative standing-wave current envelope $I(z)$ (equivalently, a volume current density $\bm{J}(\bm{r})$ localized to the metal). The key physical point is that the antenna does work on the \emph{total} field at the source, and that field is naturally decomposed into a direct component produced in the reference environment and a scattered component generated by the surroundings: $\bm{E}=\bm{E}_0+\bm{E}_{\mathrm{sc}}$. Panel~(b) emphasizes the corresponding operator viewpoint: the environment modifies the dyadic Green function $\bm{G}=\bm{G}_0+\bm{G}_{\mathrm{sc}}$, which maps currents to fields, fields to reaction power, and reaction power to measurable port quantities such as $Z_{\mathrm{in}}$ and $R_{\mathrm{in}}=R_{\mathrm{rad}}+R_{\mathrm{loss}}$. This subsection makes that mapping explicit and isolates the single quantity that will serve as the ``self'' ingredient in classical Purcell-like factors: the environment-induced change of the real input resistance, $\Delta R_{\mathrm{in}}$.

We adopt the $e^{j\omega t}$ phasor convention. For a time-harmonic current density $\bm{J}(\bm{r})e^{j\omega t}$ confined to an antenna volume $V$, the complex power delivered by the source to the fields may be written in the reaction (complex-power) form \cite{harrington1961,jackson1998,chew1995,rumsey1954}:
\begin{equation}
P_\mathrm{in} \;=\; \frac{1}{2}\int_V \bm{J}^*(\bm{r})\cdot \bm{E}(\bm{r})\, dV,
\qquad
P_\mathrm{avg}=\Re\{P_\mathrm{in}\}.
\label{eq:Pin}
\end{equation}
The total electric field at the source is related to the current distribution through the electric dyadic Green function of the \emph{full environment} (boundaries, scatterers, stratified media, etc.) \cite{chew1995,tai1994}:
\begin{equation}
\bm{E}(\bm{r}) \;=\; j\omega\mu \int_V \bm{G}(\bm{r},\bm{r}';\omega)\cdot \bm{J}(\bm{r}')\, dV',
\qquad
\bm{G}=\bm{G}_0+\bm{G}_{\mathrm{sc}}.
\label{eq:EfromG}
\end{equation}
The decomposition $\bm{G}=\bm{G}_0+\bm{G}_{\mathrm{sc}}$ (and the induced decomposition $\bm{E}=\bm{E}_0+\bm{E}_{\mathrm{sc}}$) is not an approximation: it is simply a bookkeeping choice that separates a reference problem (often free space) from the contribution due to the environment, Fig.~\ref{fig:fig2}(a). Using \eqref{eq:EfromG} in \eqref{eq:Pin} yields a quadratic functional of the current distribution:
\begin{equation}
P_\mathrm{in}
=\frac{j\omega\mu}{2}\iint_{V\times V}
\bm{J}^*(\bm{r})\cdot \bm{G}(\bm{r},\bm{r}')\cdot \bm{J}(\bm{r}')\, dV\, dV',
\label{eq:Pin_quadratic}
\end{equation}
so that the time-averaged delivered power is governed by the dissipative part of the Green operator,
\begin{equation}
P_\mathrm{avg}
=\frac{\omega\mu}{2}\iint_{V\times V}
\bm{J}^*(\bm{r})\cdot \Im\!\left\{\bm{G}(\bm{r},\bm{r}')\right\}\cdot \bm{J}(\bm{r}')\, dV\, dV'.
\label{eq:Pavg_ImG}
\end{equation}
Equation~\eqref{eq:Pavg_ImG} makes the ``Purcell-first'' content explicit: \emph{changing the environment changes $\Im\{\bm{G}\}$, hence changes the dissipative channels available to the source current, hence changes radiative damping and/or absorption.} In a lossless re-radiating environment, the change primarily corresponds to altered radiation damping. In a lossy environment, $\Im\{\bm{G}\}$ also encodes nonradiative channels (absorption); correspondingly,
\begin{equation}
P_\mathrm{avg}=P_\mathrm{rad}+P_\mathrm{abs},\qquad
P_\mathrm{abs}=\frac{1}{2}\int_{V_\mathrm{loss}}
\Big(\sigma|\bm{E}|^2+\omega\epsilon''|\bm{E}|^2+\omega\mu''|\bm{H}|^2\Big)\,dV,
\label{eq:Pabs}
\end{equation}
where $V_\mathrm{loss}$ is the region containing conductive and/or lossy dielectric/magnetic media \cite{harrington1961,jackson1998}. This is precisely why a measured increase of $R_{\mathrm{in}}$ near a lossy body cannot be interpreted as ``radiative enhancement'' unless $P_\mathrm{abs}$ (or equivalently $R_{\mathrm{loss}}$) is constrained or separately estimated.

To expose the environment-induced contribution explicitly, combine \eqref{eq:EfromG} with the Green decomposition to write
\begin{equation}
P_\mathrm{in}=P_0+P_\mathrm{env},
\qquad
P_\mathrm{env}=\frac{1}{2}\int_V \bm{J}^*\cdot \bm{E}_{\mathrm{sc}}\,dV
=\frac{j\omega\mu}{2}\iint_{V\times V}\bm{J}^*\cdot \bm{G}_{\mathrm{sc}}\cdot \bm{J}\,dV\,dV',
\label{eq:Penv}
\end{equation}
so that $\Re\{P_\mathrm{env}\}$ quantifies how the environment modifies the average power drawn by the current distribution (Fig.~\ref{fig:fig2}b).

We now connect the field-theoretic statement to port observables. For a one-port antenna with port voltage $V$ and current $I$ (defined at the feed),
\begin{equation}
Z_{\mathrm{in}}\equiv \frac{V}{I}=R_{\mathrm{in}}+jX_{\mathrm{in}},
\qquad
P_\mathrm{avg}=\frac{1}{2}\Re\{VI^*\}=\frac{1}{2}|I|^2R_{\mathrm{in}},
\qquad
R_{\mathrm{in}}=R_{\mathrm{rad}}+R_{\mathrm{loss}}.
\label{eq:port_power_impedance}
\end{equation}
Therefore, under a fixed-port-current normalization, the environment-induced change in real input resistance is
\begin{equation}
\Delta R_{\mathrm{in}}
=\frac{2\,\Re\{P_\mathrm{env}\}}{|I|^2}
=\frac{\omega\mu}{|I|^2}\iint_{V\times V}
\bm{J}^*(\bm{r})\cdot \Im\!\left\{\bm{G}_{\mathrm{sc}}(\bm{r},\bm{r}')\right\}\cdot \bm{J}(\bm{r}')\, dV\, dV',
\label{eq:DeltaRin_ImGsc}
\end{equation}
which is the precise classical statement of environment-modified damping: the environment changes $R_{\mathrm{in}}$ because it changes the available radiative/absorptive channels (encoded by $\Im\{\bm{G}_{\mathrm{sc}}\}$). A useful way to interpret \eqref{eq:DeltaRin_ImGsc} is to factor out the port current by writing $\bm{J}(\bm{r})=I\,\bm{j}(\bm{r})$, where $\bm{j}$ is a \emph{unit-current} distribution associated with the driven structure. Then
\begin{equation}
R_{\mathrm{in}}
=\omega\mu\iint_{V\times V}
\bm{j}^*(\bm{r})\cdot \Im\!\left\{\bm{G}(\bm{r},\bm{r}')\right\}\cdot \bm{j}(\bm{r}')\, dV\, dV',
\label{eq:Rin_operator}
\end{equation}
which highlights that $R_{\mathrm{in}}$ is an operator-defined overlap between the current mode supported by the antenna and the dissipative part of the environmental Green function. Importantly, the \emph{exact} current distribution $\bm{J}$ can itself change with the environment (through altered boundary conditions and loading); \eqref{eq:Pin_quadratic}--\eqref{eq:Rin_operator} remain valid because they are written for the actual $\bm{J}$ that solves the driven problem.

Finally, Fig.~\ref{fig:fig2}(b) also previews the \emph{transfer} counterpart needed for link-level effects. For two antennas with currents $\bm{J}_1$ and $\bm{J}_2$ and port currents $I_1$ and $I_2$, the mutual impedance can be written in a reaction-theorem form \cite{richmond1961,wang1975tap}:
\begin{equation}
Z_{21}
=\frac{1}{I_1 I_2}\int_{V_2}\bm{J}_2^\ast(\bm{r})\cdot \bm{E}_1(\bm{r})\,dV
=\frac{j\omega\mu}{I_1 I_2}\iint_{V_2\times V_1}
\bm{J}_2^\ast(\bm{r})\cdot \bm{G}(\bm{r},\bm{r}')\cdot \bm{J}_1(\bm{r}')\, dV\, dV',
\label{eq:Z21_G}
\end{equation}
explicitly showing that the same Green operator that governs self-damping also governs transfer/coupling via its \emph{cross} dependence on $(\bm{r},\bm{r}')$. In the next subsection, $\Delta R_{\mathrm{in}}$ will become the central ``self'' ingredient of classical Purcell-like factors, while $\bm{G}(\bm{r}_r,\bm{r}_t)$ (or $Z_{21}$) will serve as the complementary ``transfer'' ingredient used to classify link-level environmental enhancements.


\subsection{Classical Purcell-like factors: self (radiative damping) and transfer (channel coupling)}
A ``Purcell factor'' is fundamentally a statement about how efficiently electromagnetic channels remove energy from (or deliver energy to) a source. Section~2 made this precise for classical antennas: the environment enters through the dyadic Green function and changes the \emph{dissipative} part of the source self-interaction, which appears at the feed as a change in the real input impedance
$R_{\mathrm{in}}=R_{\mathrm{rad}}+R_{\mathrm{loss}}$.
This immediately suggests that any useful classical Purcell-like metric must (i) specify an \emph{excitation convention} and (ii) distinguish radiative from absorptive channels when losses are present.

\paragraph*{Self (local) factor under fixed port current.}
Under a fixed-port-current convention, the radiated power is
\begin{equation}
P_\mathrm{rad}=\frac{1}{2}|I|^2R_\mathrm{rad},
\qquad
P_\mathrm{loss}=\frac{1}{2}|I|^2R_\mathrm{loss},
\qquad
P_\mathrm{avg}=\frac{1}{2}|I|^2R_\mathrm{in}.
\label{eq:Prad_Ploss}
\end{equation}
The corresponding \emph{self} Purcell-like factor is therefore most cleanly defined in terms of the \emph{radiative} resistance,
\begin{equation}
F_{\mathrm{self}}^{(I)}
\equiv
\frac{P_\mathrm{rad}}{P_{\mathrm{rad},0}}
=
\frac{R_\mathrm{rad}}{R_{\mathrm{rad},0}}.
\label{eq:Fself_I}
\end{equation}
This definition is robust: it remains meaningful even when the environment introduces absorption, because it explicitly refers to the radiative channel rather than to the total power drawn. In experiments, however, $R_\mathrm{rad}$ is not always directly accessible, while $R_\mathrm{in}$ is readily obtained from $S_{11}\rightarrow Z_\mathrm{in}$. A useful identity makes the required separation explicit:
\begin{equation}
F_{\mathrm{self}}^{(I)}
=\frac{R_\mathrm{rad}}{R_{\mathrm{rad},0}}
=\frac{\eta_\mathrm{rad}R_\mathrm{in}}{\eta_{\mathrm{rad},0}R_{\mathrm{in},0}},
\qquad
\eta_\mathrm{rad}\equiv\frac{R_\mathrm{rad}}{R_\mathrm{in}}.
\label{eq:Fself_eta}
\end{equation}
Thus, the common impedance proxy
\begin{equation}
F_{\mathrm{self}}^{(I)}\approx \frac{R_\mathrm{in}}{R_{\mathrm{in},0}}
\label{eq:Fself_proxy}
\end{equation}
is justified only when $\eta_\mathrm{rad}$ is unchanged (or when $\Delta R_\mathrm{loss}\approx 0$) \emph{or} when $\eta_\mathrm{rad}$ has been independently measured. This caveat is exactly the Purcell-first interpretation of ``body loss'' and other proximity-loading phenomena: $R_\mathrm{in}$ may increase while $R_\mathrm{rad}$ decreases if the environment primarily adds dissipative channels. Classical radiation-shield approaches (e.g., Wheeler-type caps) and modern chamber methods provide practical routes to estimate $\eta_\mathrm{rad}$ and thus recover $R_\mathrm{rad}$ from $R_\mathrm{in}$ when needed \cite{wheeler1959}.

It is often helpful to report the \emph{loss-channel} analogue alongside the self factor,
\begin{equation}
F_{\mathrm{loss}}^{(I)}\equiv\frac{R_\mathrm{loss}}{R_{\mathrm{loss},0}},
\label{eq:Floss_I}
\end{equation}
because $F_{\mathrm{self}}^{(I)}$ and $F_{\mathrm{loss}}^{(I)}$ together fully characterize how the environment reshapes the dissipative partition of the antenna at fixed current.

\paragraph*{Self factor under a realistic source (matching matters).}
Many practical transmitters are better modeled by a Thevenin source with voltage $V_s$ and source impedance $Z_s$, so the port current depends on the environment through $Z_\mathrm{in}$.
A convenient operational expression uses the available power of the source,
$P_{\mathrm{av}}=|V_s|^2/(8\Re\{Z_s\})$ under standard assumptions, and the power-delivery factor $(1-|\Gamma|^2)$ for the mismatch between source and load \cite{pozar2011}:
\begin{equation}
P_\mathrm{del}=P_{\mathrm{av}}\,(1-|\Gamma|^2),
\qquad
\Gamma=\frac{Z_\mathrm{in}-Z_s^\ast}{Z_\mathrm{in}+Z_s}.
\label{eq:Pdel_Gamma}
\end{equation}
Since $P_\mathrm{rad}=\eta_\mathrm{rad}P_\mathrm{del}$, an operational self enhancement at fixed $P_{\mathrm{av}}$ becomes
\begin{equation}
F_{\mathrm{self}}^{(P_{\mathrm{av}})}
\equiv
\frac{P_\mathrm{rad}}{P_{\mathrm{rad},0}}
=
\frac{\eta_\mathrm{rad}}{\eta_{\mathrm{rad},0}}
\,
\frac{1-|\Gamma|^2}{1-|\Gamma_0|^2}.
\label{eq:Fself_Pav}
\end{equation}
Equation~\eqref{eq:Fself_Pav} is the practical ``RF translation'' of Purcell enhancement: the environment can increase radiated power either by increasing true radiative damping (higher $\eta_\mathrm{rad}$ at a given accepted power) \emph{or} by improving matching (higher delivered power), and both effects often occur simultaneously.

The same physics is commonly packaged in antenna-system language via \emph{realized gain}. IEEE antenna definitions distinguish gain (excluding mismatch) from realized gain (including mismatch) \cite{ieee1452013}. A convenient decomposition is
\begin{equation}
G_{\mathrm{real}}=(1-|\Gamma|^2)\,G
=(1-|\Gamma|^2)\,\eta_\mathrm{rad}\,D,
\label{eq:Greall}
\end{equation}
where $D$ is directivity. This factorization will be used repeatedly in the case studies: the environment can alter $R_\mathrm{rad}$ (damping), $R_\mathrm{loss}$ (absorption), matching $(1-|\Gamma|^2)$, and/or $D$ (channel partition into directions/polarizations). The Purcell-first hierarchy remains: changes in $R_\mathrm{rad}$ (and thus $\eta_\mathrm{rad}$) are the \emph{primary} signature of local Purcell-like enhancement; pattern changes are the \emph{secondary} signature of how the environment redistributes the radiated power once the coupling to channels has changed.

\paragraph*{Equivalence to LDOS/Green-tensor formulas (small-emitter limit).}
For electrically small dipole-like emitters, the impedance and Green-function viewpoints become explicitly identical. For an electric dipole moment $\bm{p}$ at $\bm{r}_0$,
\begin{equation}
P_\mathrm{rad}(\bm{r}_0)
=\frac{\omega}{2}\Im\!\left\{\bm{p}^*\cdot \bm{E}(\bm{r}_0)\right\}
=\frac{\omega^3\mu}{2}\,\bm{p}^*\cdot \Im\!\{\bm{G}(\bm{r}_0,\bm{r}_0)\}\cdot \bm{p},
\label{eq:Pdipole_G}
\end{equation}
so that
\begin{equation}
F_P(\bm{r}_0)
=
\frac{\bm{p}^*\cdot \Im\{\bm{G}(\bm{r}_0,\bm{r}_0)\}\cdot \bm{p}}
     {\bm{p}^*\cdot \Im\{\bm{G}_0(\bm{r}_0,\bm{r}_0)\}\cdot \bm{p}}.
\label{eq:FP_G}
\end{equation}


For an electric point dipole with complex dipole moment $\bm{p}=p\,\hat{\bm{u}}$
located at $\bm{r}_0$ and oscillating at angular frequency $\omega$,
the environment-modified (total) decay rate $\Gamma$ (or, equivalently, the time-averaged power dissipated by the dipole)
is governed by the on-site dyadic Green tensor via the projected contraction
$\hat{\bm{u}}\!\cdot\!\Im\{\bm{G}(\bm{r}_0,\bm{r}_0;\omega)\}\!\cdot\!\hat{\bm{u}}$.
A convenient orientation-resolved Purcell factor is therefore written as the ratio
\begin{equation}
F_P(\bm{r}_0,\omega;\hat{\bm{u}})
\;\equiv\;\frac{\Gamma}{\Gamma_0}
\;=\;
\frac{\hat{\bm{u}}\cdot\Im\!\left\{\bm{G}(\bm{r}_0,\bm{r}_0;\omega)\right\}\cdot\hat{\bm{u}}}
     {\hat{\bm{u}}\cdot\Im\!\left\{\bm{G}_0(\bm{r}_0,\bm{r}_0;\omega)\right\}\cdot\hat{\bm{u}}}\,,
\label{eq:FP_projected}
\end{equation}
where $\bm{G}$ is the full Green tensor of the actual environment, $\bm{G}_0$ is the Green tensor of the chosen reference
(background) environment (e.g., homogeneous free space or a homogeneous host medium), and $\hat{\bm{u}}$ is a unit vector
that specifies the dipole orientation.  For an \emph{isotropically oriented} dipole (or orientation-averaged emission),
$\langle \hat{\bm{u}}\,\hat{\bm{u}} \rangle=(1/3)\bm{I}$ and the contraction reduces to a trace:
\begin{equation}
\overline{F}_P(\bm{r}_0,\omega)
\;=\;\frac{1}{3}\frac{\Tr\,\Im\{\bm{G}(\bm{r}_0,\bm{r}_0;\omega)\}}
                       {\,\frac{1}{3}\Tr\,\Im\{\bm{G}_0(\bm{r}_0,\bm{r}_0;\omega)\}}
\;=\;
\frac{\Tr\,\Im\{\bm{G}(\bm{r}_0,\bm{r}_0;\omega)\}}{\Tr\,\Im\{\bm{G}_0(\bm{r}_0,\bm{r}_0;\omega)\}}\,.
\label{eq:FP_isotropic}
\end{equation}
Since the (electric) LDOS is proportional to $\Im[\Tr\,\bm{G}(\bm{r}_0,\bm{r}_0;\omega)]$, Eq.~\eqref{eq:FP_isotropic}
makes explicit that the \emph{self} enhancement for an isotropic dipole is the LDOS enhancement (up to the chosen normalization).
This also clarifies a frequent experimental caveat: $F_P$ is a \emph{dissipative-channel} enhancement (radiative plus absorptive);
in lossy environments a large $F_P$ can correspond to increased nonradiative dissipation (“quenching”) rather than increased far-field radiation.

The commonly used single-mode cavity estimate,
\begin{equation}
F_{P,\mathrm{cav}} \approx \frac{3}{4\pi^2}\left(\frac{\lambda}{n}\right)^3\frac{Q}{V_\mathrm{eff}},
\label{eq:Purcell_QV}
\end{equation}
should be interpreted as the \emph{peak} (on-resonance) enhancement under restrictive assumptions:
(i) a dominant \emph{single} cavity-like resonance with approximately Lorentzian response,
(ii) weak coupling (no vacuum Rabi splitting), and
(iii) a well-defined energy-normalized mode profile that permits a meaningful mode volume.
Here $\lambda=2\pi c/\omega_0$ is the free-space wavelength at the resonant frequency $\omega_0$,
$n$ is the refractive index at the emitter location, $Q\equiv \omega_0/(2\gamma)$ is the quality factor
(with $\gamma$ the modal energy-decay rate), and $V_\mathrm{eff}$ is an effective mode volume.
A more explicit single-resonance expression highlights the usually implicit spatial, polarization, and detuning factors:
\begin{equation}
F_{P,\mathrm{1m}}(\bm{r}_0,\omega;\hat{\bm{u}})
\;\approx\;
\frac{3}{4\pi^2}\left(\frac{\lambda}{n}\right)^3\frac{Q}{V_\mathrm{eff}}
\;\times\;
\underbrace{\frac{|\hat{\bm{u}}\cdot \bm{E}(\bm{r}_0)|^2}{\max_{\bm{r}}|\bm{E}(\bm{r})|^2}}_{\text{polarization/spatial overlap}}
\;\times\;
\underbrace{\frac{1}{1+4Q^2\left(\frac{\omega-\omega_0}{\omega_0}\right)^2}}_{\text{spectral detuning}}\,,
\label{eq:Purcell_1mode_overlap}
\end{equation}
where $\bm{E}(\bm{r})$ is the (appropriately normalized) modal electric field.
Equation~\eqref{eq:Purcell_1mode_overlap} makes clear why naive $Q/V_\mathrm{eff}$ heuristics can fail away from the field maximum
or for emitters with unfavorable orientation.

For open, strongly radiating, strongly absorbing, and/or dispersive resonant structures (typical in plasmonics and in many RF
installations), $Q$ and $V_\mathrm{eff}$ can become ambiguous (and, in rigorous treatments, $V_\mathrm{eff}$ may be complex and
frequency dependent).  In these regimes, quantitatively reliable evaluation of $F_P$ should revert to the Green-tensor/LDOS
definition in Eqs.~\eqref{eq:FP_projected}--\eqref{eq:FP_isotropic} or to quasinormal-mode (QNM) generalizations that provide a
proper modal expansion and normalization for leaky resonators, including non-Lorentzian spectral behavior in the presence of
dissipation \cite{koenderink2010ol,sauvan2013prl,kristensen2015pra}.

\paragraph*{Transfer (link-level) factor: coupling through the cross Green function (or mutual impedance).}
Many ``real-life'' range changes are primarily \emph{transfer} effects: the environment modifies coupling between a transmitter (Tx) at $\bm{r}_t$ and a receiver (Rx) at $\bm{r}_r$ by reshaping the \emph{cross} Green function $\bm{G}(\bm{r}_r,\bm{r}_t)$, as previewed in Fig.~2(b). In circuit terms, the same physics appears as environment-dependent mutual impedance $Z_{21}$ (Section~2). A particularly transparent receive-side interpretation uses the open-circuit voltage induced at the Rx port (with $I_2=0$):
\begin{equation}
V_{\mathrm{oc},2}=V_2\big|_{I_2=0}=Z_{21}I_1,
\label{eq:Voc_Z21}
\end{equation}
so that the \emph{maximum available} received power under conjugate matching at the receiver is
\begin{equation}
P_{\mathrm{av},2}=\frac{|V_{\mathrm{oc},2}|^2}{8\,\Re\{Z_{\mathrm{in},2}\}}
=\frac{|Z_{21}|^2\,|I_1|^2}{8\,\Re\{Z_{\mathrm{in},2}\}}.
\label{eq:Pav2_Z21}
\end{equation}
When the Rx antenna and termination are unchanged, a natural transfer enhancement proxy is therefore $|Z_{21}|^2$ (or $|S_{21}|^2$ under fixed port conditions); more generally, one can normalize by $\Re\{Z_{\mathrm{in},2}\}$ to remove receiver-side impedance changes.

At the dipole level, the same transfer physics can be expressed directly through the cross Green tensor contraction:
\begin{equation}
F_{\mathrm{tr}}
\equiv
\frac{\left|\bm{p}_r^\dagger\,\bm{G}(\bm{r}_r,\bm{r}_t;\omega)\,\bm{p}_t\right|^2}
     {\left|\bm{p}_r^\dagger\,\bm{G}_0(\bm{r}_r,\bm{r}_t;\omega)\,\bm{p}_t\right|^2},
\label{eq:Ftr}
\end{equation}
which makes explicit that transfer enhancement depends not only on geometry but also on polarization/mode matching (through $\bm{p}_t$ and $\bm{p}_r$). The same cross-Green object underlies the cross density of states (CDOS), a coherence-based measure of how strongly two points are coupled through the electromagnetic environment \cite{caze2013prl}:
\begin{equation}
\rho(\bm{r},\bm{r}';\omega)=\frac{2\omega}{\pi c^2}\,\Im\!\left[\Tr\,\bm{G}(\bm{r},\bm{r}';\omega)\right].
\label{eq:CDOS}
\end{equation}

Taken together, $F_{\mathrm{self}}$ and $F_{\mathrm{tr}}$ provide a minimal but complete classical framework: \emph{self} factors quantify environment-induced changes in local radiative damping (the partition $R_\mathrm{rad}$ versus $R_\mathrm{loss}$, plus matching under a realistic source), while \emph{transfer} factors quantify environment-induced changes in channel coupling between two locations (cross Green function / mutual impedance). The taxonomy in the next section uses this distinction to classify real-life ``range boosts'' by their dominant physical lever and by the most diagnostic measurable.


\subsection{From Purcell-like modification to link budgets: Friis, two-ray, and range scalings}
Link budgets are where the ``self vs.\ transfer'' distinction becomes operational. A convenient way to expose that structure is to write the received power in the form
\begin{equation}
P_r(\omega)=P_t(\omega)\,G_{t,\mathrm{real}}(\omega)\,G_{r,\mathrm{real}}(\omega)\,|h(\omega)|^2\,L_\mathrm{sys}^{-1},
\label{eq:Pr_general}
\end{equation}
where $G_{t,\mathrm{real}}$ and $G_{r,\mathrm{real}}$ are the realized gains (including mismatch and radiation efficiency) defined in \eqref{eq:Greall}, $L_\mathrm{sys}$ groups implementation losses (cables, connector loss, polarization mismatch, etc.), and $h(\omega)$ is a scalar channel coefficient for the chosen Tx/Rx polarizations (or, more generally, a matrix coefficient for MIMO). In the Green-function language of Section~2, $h$ is the link-level manifestation of the \emph{cross} interaction $\bm{G}(\bm{r}_r,\bm{r}_t)$; therefore any environment-induced change of $|h|^2$ is naturally a \emph{transfer} Purcell-like effect, while any environment-induced change of $G_{\mathrm{real}}$ is a \emph{self} effect.

\textbf{Free-space (Friis) as the transfer baseline.}
In the far field of both antennas, in a homogeneous medium, and for co-polarized alignment, the channel coefficient reduces to the familiar spherical-wave factor
\begin{equation}
h_0(\omega)=\frac{\lambda}{4\pi R}\,e^{-jkR},
\qquad
k=\frac{2\pi}{\lambda},
\label{eq:h0}
\end{equation}
which yields the Friis transmission formula \cite{friis1946,balanis2016}
\begin{equation}
P_r \;=\; P_t\, G_{t,\mathrm{real}}\, G_{r,\mathrm{real}}
\left(\frac{\lambda}{4\pi R}\right)^2\,L_\mathrm{sys}^{-1}.
\label{eq:Friis}
\end{equation}
Recommendation ITU-R P.525 defines the associated free-space basic transmission loss,
\begin{equation}
L_\mathrm{bf}=20\log_{10}\!\left(\frac{4\pi R}{\lambda}\right)\ \mathrm{dB},
\label{eq:fspl_itu}
\end{equation}
and provides practical forms in $(f,d)$ units \cite{itu525}. In this baseline regime, the transfer factor is unity by definition. It is therefore natural to define a \emph{transfer Purcell-like factor} as a normalized channel enhancement
\begin{equation}
F_{\mathrm{tr}}(\omega)\equiv\frac{|h(\omega)|^2}{|h_0(\omega)|^2},
\label{eq:Ftr_h}
\end{equation}
so that \eqref{eq:Pr_general} becomes $P_r=P_t\,G_{t,\mathrm{real}}G_{r,\mathrm{real}}(\lambda/4\pi R)^2F_{\mathrm{tr}}\,L_\mathrm{sys}^{-1}$. This matches the dipole-level $F_{\mathrm{tr}}$ introduced earlier (Section~2) but is expressed in a directly measurable link-budget form.

A practical range calculation is set by a sensitivity threshold $P_r\ge P_{\min}$. A standard receiver model writes
\begin{equation}
P_{\min}=N + \mathrm{SNR}_{\min},
\qquad
N=kT\,B\,F,
\label{eq:Pmin_noise}
\end{equation}
where $B$ is bandwidth and $F$ is the receiver noise factor (noise figure $\mathrm{NF}=10\log_{10}F$). In dBm units this becomes the familiar
\begin{equation}
N\,[\mathrm{dBm}]\approx -174 + 10\log_{10}B[\mathrm{Hz}] + \mathrm{NF}\,[\mathrm{dB}],
\label{eq:noise_dBm}
\end{equation}
and the link margin is $M=P_r-P_{\min}$ \cite{rappaport2002,goldsmith2005}. Substituting \eqref{eq:Friis} into $P_r=P_{\min}$ yields the free-space range scaling
\begin{equation}
\frac{R'}{R_0}=\sqrt{\frac{G'_{t,\mathrm{real}}}{G_{t,\mathrm{real}}}}
=10^{\Delta G_{t,\mathrm{real}}/20},
\label{eq:Rscale_fs}
\end{equation}
demonstrating that any \emph{self} mechanism that increases realized gain (via $\eta_\mathrm{rad}$, matching, or directivity) increases range proportionally to the square root in the $R^{-2}$ regime.

\textbf{Two-ray (ground-bounce) as a transfer-dominated canonical example.}
Near a dominant planar boundary (ground, sea, runway, large conducting wall), the received field is often governed by the coherent sum of a direct path and a reflected path. A deterministic two-ray model writes the received power as \cite{jakes1974,parsons2000,itu1411}
\begin{equation}
P_r
=
P_t\, G_{t,\mathrm{real}}\,G_{r,\mathrm{real}}
\left(\frac{\lambda}{4\pi}\right)^2
\left|
\frac{e^{-jkR_1}}{R_1}
+\Gamma(\theta_i)\,\frac{e^{-jkR_2}}{R_2}
\right|^2
L_\mathrm{sys}^{-1},
\label{eq:two_ray_full}
\end{equation}
where $R_1=\sqrt{R^2+(h_t-h_r)^2}$ and $R_2=\sqrt{R^2+(h_t+h_r)^2}$ are the direct and reflected path lengths for antenna heights $h_t,h_r$, and $\Gamma(\theta_i)$ is the Fresnel reflection coefficient at incidence angle $\theta_i$ (with polarization dependence and complex permittivity for lossy ground). For a planar interface with relative permittivity $\tilde{\epsilon}_r$ (including conductivity), the Fresnel coefficients are \cite{jackson1998,balanis2016}
\begin{equation}
\Gamma_\perp(\theta_i)=\frac{\cos\theta_i-\sqrt{\tilde{\epsilon}_r-\sin^2\theta_i}}
{\cos\theta_i+\sqrt{\tilde{\epsilon}_r-\sin^2\theta_i}},
\qquad
\Gamma_\parallel(\theta_i)=\frac{\tilde{\epsilon}_r\cos\theta_i-\sqrt{\tilde{\epsilon}_r-\sin^2\theta_i}}
{\tilde{\epsilon}_r\cos\theta_i+\sqrt{\tilde{\epsilon}_r-\sin^2\theta_i}}.
\label{eq:fresnel}
\end{equation}
Equation~\eqref{eq:two_ray_full} is an explicit \emph{transfer} factor: it modifies the channel coefficient $h$ through an additional coherent path. Indeed, relative to the free-space baseline, one may write the two-ray transfer enhancement as
\begin{equation}
F_{\mathrm{tr}}^{(2\mathrm{ray})}
=
\left|
1+\Gamma(\theta_i)\,\frac{R_1}{R_2}\,e^{-jk(R_2-R_1)}
\right|^2,
\label{eq:Ftr_2ray}
\end{equation}
showing directly why modest geometry changes can produce deep nulls or large constructive boosts even when $G_{t,\mathrm{real}}$ and $G_{r,\mathrm{real}}$ are unchanged.

In the far two-ray regime ($R\gg h_t,h_r$), the path difference admits the approximation $R_2-R_1\approx 2h_th_r/R$ and, for many grounds at low grazing angles, $|\Gamma|\approx 1$ with a phase close to $\pi$. In that regime the envelope approaches the well-known $R^{-4}$ law (often used as a lower bound in ITU-R P.1411) \cite{itu1411}:
\begin{equation}
P_r \;\propto\;
P_t\,G_{t,\mathrm{real}}\,G_{r,\mathrm{real}}
\left(\frac{h_t h_r}{R^2}\right)^2
L_\mathrm{sys}^{-1}.
\label{eq:two_ray_asymptote}
\end{equation}
The corresponding breakpoint distance is approximately \cite{itu1411}
\begin{equation}
R_\mathrm{bp}\approx \frac{4h_t h_r}{\lambda},
\label{eq:Rbp_itu}
\end{equation}
and ITU-R P.1411 explicitly uses two slopes (20 and 40 dB/decade) with a fading margin for practical bounds \cite{itu1411}. A key practical consequence is that gain-driven range improvement is weaker in the $R^{-4}$ regime:
\begin{equation}
\frac{R'}{R_0}
=\left(\frac{G'_{t,\mathrm{real}}}{G_{t,\mathrm{real}}}\right)^{1/4}
=10^{\Delta G_{t,\mathrm{real}}/40},
\qquad (R\gg R_\mathrm{bp}),
\label{eq:Rscale_2ray}
\end{equation}
so a few dB of self enhancement produces a smaller fractional range change than in free space, while transfer changes (through $F_{\mathrm{tr}}$) can still create very large swings.

\textbf{General path-loss exponents and a compact ``range scaling rule.''}
Many environments are well summarized (over a limited range) by a power-law path loss $P_r\propto R^{-n}$ with exponent $n$ (e.g., $n\approx 2$ free space, $n\approx 4$ far two-ray, intermediate values in many outdoor/indoor scenarios) \cite{rappaport2002,parsons2000}. In that case, an incremental realized-gain improvement $\Delta G$ (in dB) translates into the general scaling
\begin{equation}
\frac{R'}{R_0}=10^{\Delta G/(10n)}.
\label{eq:Rscale_n}
\end{equation}
Equation~\eqref{eq:Rscale_n} is a useful ``back-of-the-envelope'' rule for turning dB-level self improvements into range changes once the dominant distance law is known.

\textbf{Mapping back to self/transfer Purcell-like mechanisms.}
Equations~\eqref{eq:Pr_general}--\eqref{eq:Rscale_n} make the separation clear. In a fixed-geometry experiment, changes in $G_{t,\mathrm{real}}$ and $G_{r,\mathrm{real}}$ are primarily \emph{self} effects (changes in $R_\mathrm{rad}$, $R_\mathrm{loss}$, matching, and/or directivity), whereas changes in $|h|^2$ are \emph{transfer} effects (added coherent paths, waveguiding, confinement, or engineered scattering). This is the reason that ``pattern improvement'' language is often ambiguous: the same received-power change can be produced by a local self enhancement (larger $G_{\mathrm{real}}$) or by a transfer enhancement (larger $F_{\mathrm{tr}}$) without any change in the standalone antenna. Engineered environments (reflectarrays/metasurfaces/RIS) may be viewed as deliberate attempts to program the scattering contribution $\bm{G}_{\mathrm{sc}}$ to increase $F_{\mathrm{tr}}$ at selected locations; in such settings, one must also distinguish far-field scaling from near-field and aperture-limited behavior \cite{direnzo2020jsac,bjornson2020ojcoms}. The next section uses this framework to classify real-life Purcell-like effects according to which mechanism dominates, and to identify the most diagnostic measurements for each class.

\begin{figure*}[t]
    \centering
    \includegraphics[width=0.98\textwidth]{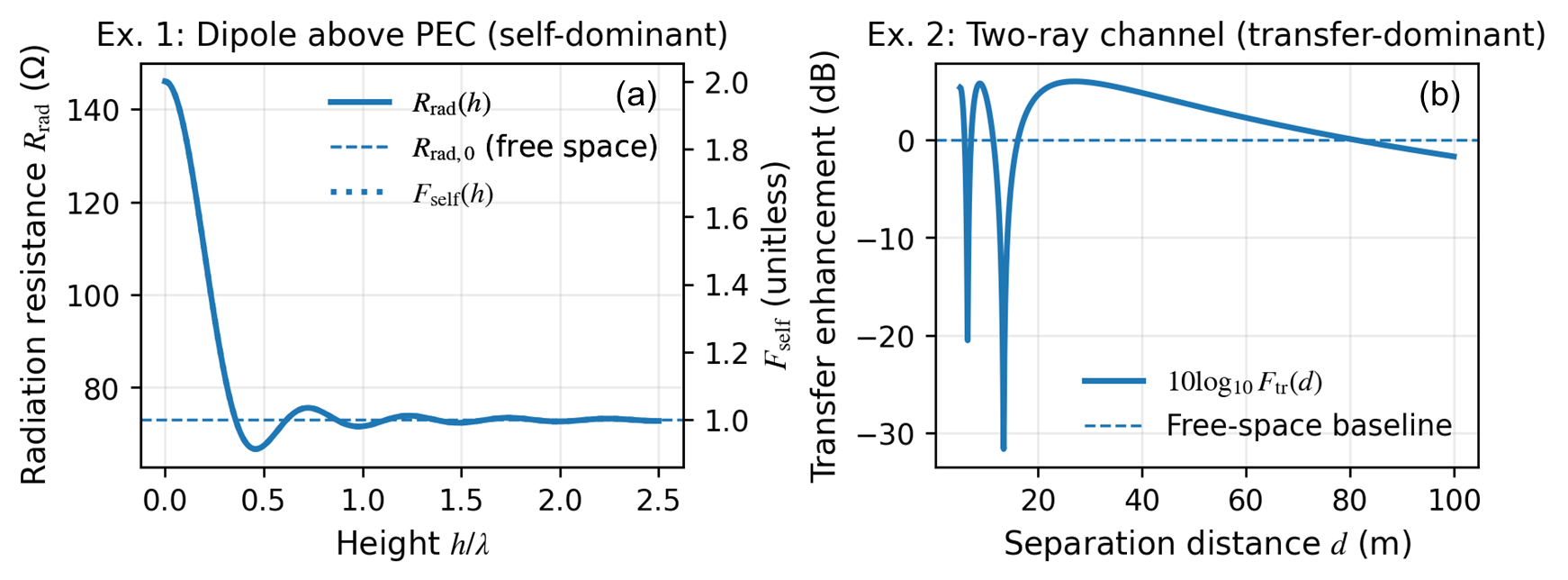}
    \caption{
    Two canonical “worked examples” that validate the self/transfer factorization end-to-end.
    \textbf{(a) Dipole above a perfect electric conductor (PEC): self-dominant.}
    A vertical radiator at height $h$ above an infinite PEC experiences an environment-induced change in radiative damping (mirror/image interference), quantified here by the radiation resistance $R_{\mathrm{rad}}(h)$ (solid).
    The dashed line shows the free-space reference $R_{\mathrm{rad},0}$.
    The dotted curve (right axis) is the self-enhancement factor
    $F_{\mathrm{self}}(h)\equiv R_{\mathrm{rad}}(h)/R_{\mathrm{rad},0}$, exhibiting Drexhage-like oscillations versus $h/\lambda$ and approaching unity at large separations.
    \textbf{(b) Two-ray channel: transfer-dominant.}
    The transfer enhancement $10\log_{10}F_{\mathrm{tr}}(d)$ (solid) versus link distance $d$ for a direct path plus a single ground-reflected path.
    Here $F_{\mathrm{tr}}(d)\equiv |h(d)|^{2}/|h_{\mathrm{fs}}(d)|^{2}$ isolates channel physics relative to the free-space baseline (dashed, 0~dB), producing deep fades and constructive regions while (by construction) the antenna port match can remain approximately unchanged.
    }
    \label{fig:worked_examples_self_transfer}
\end{figure*}

\subsection{Two canonical worked examples: separating self and transfer}
To ensure the framework is not purely conceptual, we include two canonical examples in Fig.~\ref{fig:worked_examples_self_transfer} that isolate the two factors in Eq.~\eqref{eq:Pr_ratio_factorization}, see below.
The first example is intentionally \emph{self-dominant}: the environment primarily modifies the radiative damping (and hence the radiation resistance for fixed current), while the second is \emph{transfer-dominant}: the propagation channel varies strongly while the antenna’s local port properties can remain essentially unchanged.

\paragraph{Example 1: dipole above PEC (self-dominant).}
Consider a vertically oriented radiator at height $h$ above an infinite, perfectly conducting plane.
By image theory, the field in the upper half-space is equivalent to that of the physical source plus an in-phase image source below the plane.
This produces distance-dependent constructive/destructive interference in the radiated field, leading to an oscillatory radiated-power (and radiative-damping) modulation analogous to the classic Drexhage effect for dipoles near planar reflectors.
In the constant-current convention, the self-enhancement is naturally expressed as
\begin{equation}
F_{\mathrm{self}}(h)\;\equiv\;\frac{R_{\mathrm{rad}}(h)}{R_{\mathrm{rad},0}}
\;=\;\frac{P_{\mathrm{rad}}(h)}{P_{\mathrm{rad},0}},
\label{eq:Fself_def}
\end{equation}
where $R_{\mathrm{rad},0}$ and $P_{\mathrm{rad},0}$ denote the free-space reference values.
Figure~\ref{fig:worked_examples_self_transfer}(a) plots $R_{\mathrm{rad}}(h)$ together with $F_{\mathrm{self}}(h)$, showing oscillations versus $h/\lambda$ and convergence to the free-space limit for $h\gg\lambda$.
Crucially, this “self” modulation is \emph{distinct} from impedance mismatch: the latter is quantified independently through $|\Gamma|$ and enters the realized gain through the standard mismatch factor $(1-|\Gamma|^{2})$ (see definitions in IEEE Std.~145).

\paragraph{Example 2: two-ray channel (transfer-dominant).}
To isolate a pure transfer effect, we adopt a minimal two-ray ground-reflection channel with a direct path of length
$l_{1}=\sqrt{d^{2}+(h_{t}-h_{r})^{2}}$ and a reflected path of length
$l_{2}=\sqrt{d^{2}+(h_{t}+h_{r})^{2}}$, yielding the narrowband complex channel
\begin{equation}
h(d)\;=\;\frac{e^{-jkl_{1}}}{l_{1}}+\rho(\theta)\,\frac{e^{-jkl_{2}}}{l_{2}},
\label{eq:two_ray_channel}
\end{equation}
with $k=2\pi/\lambda$ and $\rho(\theta)$ the Fresnel reflection coefficient (often $\rho\approx-1$ for a highly conducting ground at grazing incidence).
We define a transfer enhancement relative to free-space propagation as
\begin{equation}
F_{\mathrm{tr}}(d)\;\equiv\;\frac{|h(d)|^{2}}{|h_{\mathrm{fs}}(d)|^{2}},
\qquad
h_{\mathrm{fs}}(d)=\frac{e^{-jkl_{1}}}{l_{1}}.
\label{eq:Ftr_def}
\end{equation}
Figure~\ref{fig:worked_examples_self_transfer}(b) shows that $F_{\mathrm{tr}}(d)$ can vary by many dB due to two-path interference (deep fades and constructive regions), even when the antenna’s local matching (and thus the self factor) is held fixed.
This illustrates why received-power “enhancement” alone cannot be interpreted as a Purcell-like self effect without independent evidence from port observables (e.g., $S_{11}$, extracted $R_{\mathrm{rad}}$, or radiation efficiency).

\section{A taxonomy of real-life Purcell-like effects}
Section~2 established the central organizing principle of this review: an ``environmental enhancement'' becomes \emph{Purcell-like} when it can be traced to an environment-induced change in the \emph{dissipative electromagnetic channels} accessible to a driven source current distribution. In the Green-function viewpoint, this change is encoded by the environment-modified dyadic Green function $\bm{G}=\bm{G}_0+\bm{G}_{\mathrm{sc}}$ and by how $\Im\{\bm{G}\}$ controls the time-averaged work done by the current (Eq.~\eqref{eq:Pavg_ImG}). In the port viewpoint, the same physics appears as a change of the real input impedance
\(
R_{\mathrm{in}}=R_{\mathrm{rad}}+R_{\mathrm{loss}}
\)
(Eq.~\eqref{eq:port_power_impedance}), i.e., a change of radiative damping and/or absorption for the coupled source--environment system (Fig.~\ref{fig:fig2}).

For practical systems, what is typically reported is not $R_{\mathrm{rad}}$ itself but a change in received power, range, or coverage. The most useful diagnostic is therefore to factorize the received-power change into a \emph{local} (self) part and a \emph{channel} (transfer) part. Using the link-budget form introduced in Section~2, the received-power ratio between an environment of interest and a reference environment can be written as
\begin{equation}
\frac{P_r}{P_{r,0}}
=
\underbrace{\frac{G_{t,\mathrm{real}}}{G_{t,\mathrm{real},0}}
\frac{G_{r,\mathrm{real}}}{G_{r,\mathrm{real},0}}}_{\text{self (local): matching $\times$ efficiency $\times$ directivity}}
\;
\underbrace{\frac{|h|^2}{|h_0|^2}}_{\text{transfer (link): channel coupling}}
=
\frac{G_{t,\mathrm{real}}}{G_{t,\mathrm{real},0}}
\frac{G_{r,\mathrm{real}}}{G_{r,\mathrm{real},0}}
\,F_{\mathrm{tr}},
\label{eq:Pr_ratio_factorization}
\end{equation}
where $G_{\mathrm{real}}=(1-|\Gamma|^2)\eta_{\mathrm{rad}}D$ (Eq.~\eqref{eq:Greall}) captures \emph{local} antenna changes (matching, efficiency via $R_{\mathrm{rad}}/(R_{\mathrm{rad}}+R_{\mathrm{loss}})$, and directionality), while $F_{\mathrm{tr}}$ captures \emph{channel} changes (additional coherent paths, guiding/confinement, or engineered scattering). Equation~\eqref{eq:Pr_ratio_factorization} is the practical backbone of the taxonomy: many ``range boosts'' are dominated by either a change in realized gain (self) or a change in channel coupling (transfer), even though both may be present.

\begin{figure*}[t]
\centering
\includegraphics[width=\textwidth]{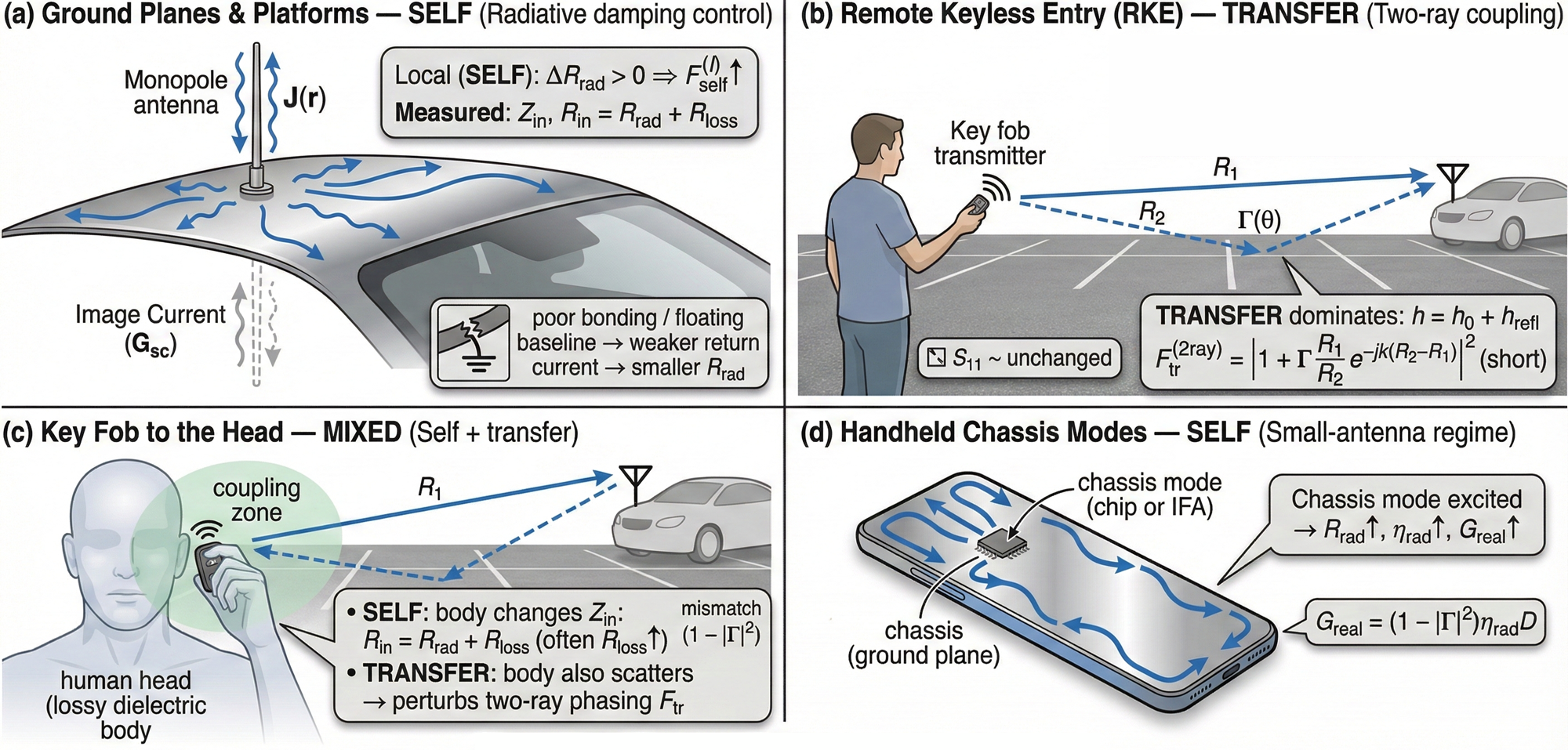}
\caption{\textbf{Everyday archetypes of self vs.\ transfer Purcell-like enhancement (schematic).}
(a) \textbf{Ground planes/platforms---SELF:} a conducting platform modifies the local self-interaction, strengthening radiative damping (often $\Delta R_{\mathrm{rad}}>0$ relative to poor return-path baselines).
(b) \textbf{RKE near-ground links---TRANSFER:} the ground introduces a second coherent path; received power is controlled by two-ray coupling (large swings in $F_{\mathrm{tr}}$ can occur even when $S_{11}$ is nearly unchanged).
(c) \textbf{``Key fob to the head''---MIXED:} body proximity perturbs $Z_{\mathrm{in}}$ and loss partition (self) while also scattering and perturbing two-ray phasing (transfer).
(d) \textbf{Handheld chassis modes---SELF:} exciting chassis currents increases $R_{\mathrm{rad}}$ and $\eta_{\mathrm{rad}}$, raising realized gain and bandwidth in the small-antenna regime.}
\label{fig:fig3}
\end{figure*}

\begin{figure*}[t]
\centering
\includegraphics[width=\textwidth]{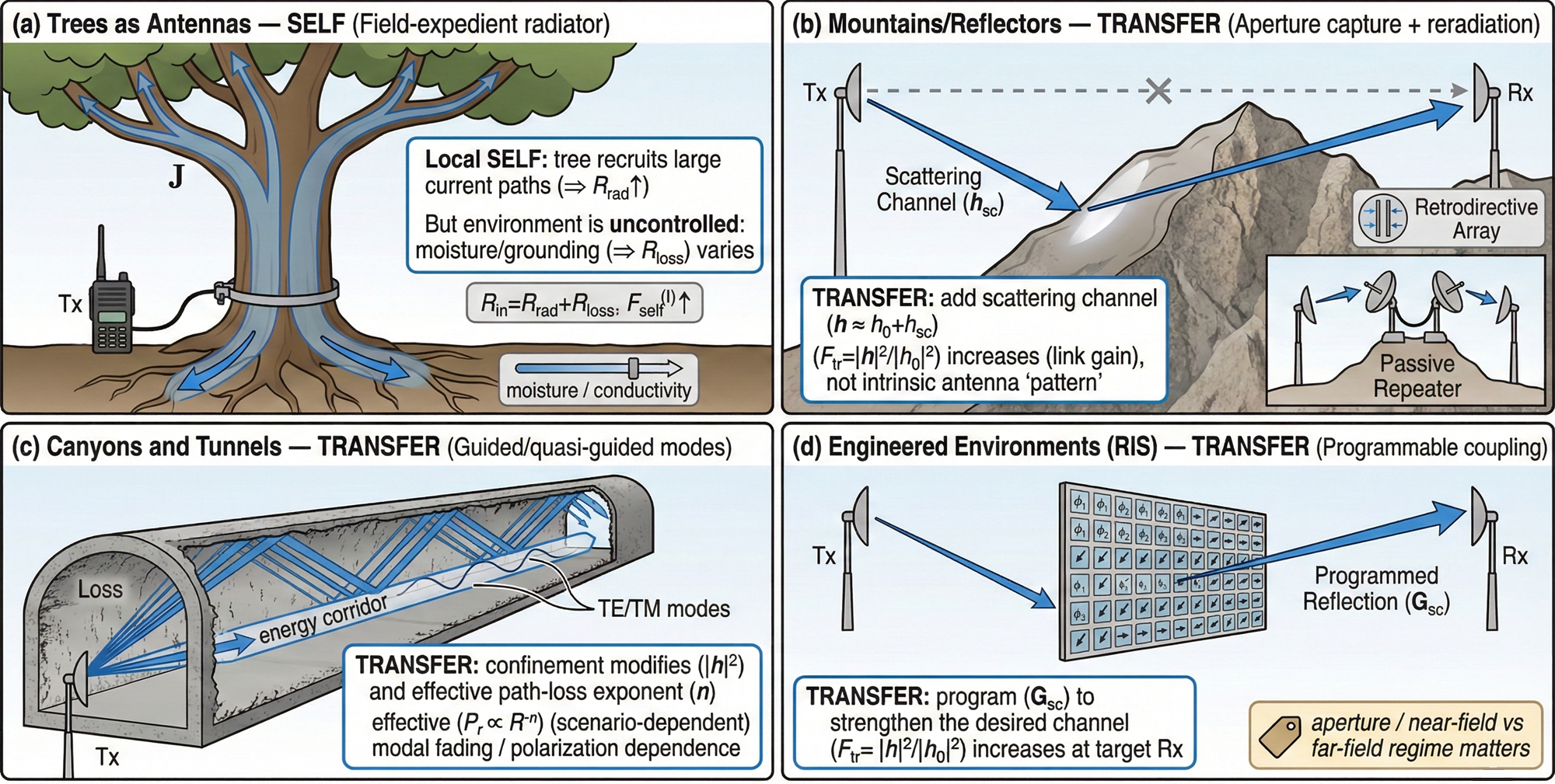}
\caption{\textbf{Environments that supply or engineer additional radiative/coupling channels (schematic).}
(a) \textbf{Trees as antennas---SELF:} an excited tree recruits large current paths (potentially $R_{\mathrm{rad}}\uparrow$) but is highly variable because moisture/grounding also change $R_{\mathrm{loss}}$.
(b) \textbf{Mountains/reflectors/passive repeaters---TRANSFER:} a reflector-induced scattering channel adds to the direct channel ($h\!\approx\!h_0+h_{\mathrm{sc}}$), increasing the link coupling $F_{\mathrm{tr}}$; retrodirective arrays provide robust angular response.
(c) \textbf{Canyons/tunnels---TRANSFER:} confinement supports guided/quasi-guided propagation; effective path-loss behavior and polarization/mode content dominate performance.
(d) \textbf{RIS/metasurfaces---TRANSFER:} a programmable scattering environment is configured to strengthen a desired channel; performance depends on aperture and near-/far-field regime.}
\label{fig:fig4}
\end{figure*}

Figures~\ref{fig:fig3} and \ref{fig:fig4} provide a pictorial taxonomy of representative scenarios. Figure~\ref{fig:fig3} emphasizes canonical \emph{self-dominant}, \emph{transfer-dominant}, and \emph{mixed} effects in everyday wireless links (platform/ground plane; RKE two-ray; ``fob-to-head''; chassis-mode excitation). Figure~\ref{fig:fig4} emphasizes environments that supply or engineer additional radiative/coupling channels (trees as field-expedient radiators; mountains/reflectors/passive repeaters; canyon/tunnel confinement; programmable RIS/metasurfaces).

This motivates two complementary Purcell-like metrics that recur throughout the case studies:

\emph{(i) Self (local) Purcell-like enhancement.}
The environment modifies the \emph{self-interaction} of a radiator through $\bm{G}(\bm{r}_t,\bm{r}_t)$, changing $R_{\mathrm{rad}}$ and possibly $R_{\mathrm{loss}}$ (local damping and absorption). Under fixed port current, the natural measure is $F_{\mathrm{self}}^{(I)}=R_{\mathrm{rad}}/R_{\mathrm{rad},0}$, while under a realistic source the operational radiated power also depends on matching (through $\Gamma$ and the delivered-power factor). In all cases, radiation-pattern changes are interpreted as \emph{secondary}: they reflect how power is redistributed among channels once the local dissipative coupling has been modified.

\emph{(ii) Transfer (link-level) Purcell-like enhancement.}
The environment modifies \emph{cross coupling} between Tx and Rx through $\bm{G}(\bm{r}_r,\bm{r}_t)$ (equivalently, mutual coupling/impedance), changing the channel coefficient $h$ and thus $F_{\mathrm{tr}}=|h|^2/|h_0|^2$. This class includes two-ray interference (Fig.~\ref{fig:fig3}b), waveguiding and confinement (Fig.~\ref{fig:fig4}c), and engineered scattering/redirection (Fig.~\ref{fig:fig4}b,d). Here ``pattern improvement'' language is often misleading: the dominant lever is a change in the \emph{channel} rather than a change in the standalone radiator.

Most real scenarios contain \emph{both} ingredients (Fig.~\ref{fig:fig3}c). The purpose of the taxonomy is therefore not to force a single label, but to identify the \emph{primary lever} and the most diagnostic observable to avoid common pitfalls---chiefly, confusing an increase in $R_{\mathrm{in}}$ with increased radiated power when $R_{\mathrm{loss}}$ also increases, or attributing a channel-induced swing in $F_{\mathrm{tr}}$ to a change in antenna gain.

\begin{table}[!t]
\centering
\caption{Taxonomy of representative classical Purcell-like effects organized by the dominant mechanism. ``SELF'' primarily changes realized gain through the local partition $R_{\mathrm{in}}=R_{\mathrm{rad}}+R_{\mathrm{loss}}$ and matching; ``TRANSFER'' primarily changes channel coupling through $F_{\mathrm{tr}}=|h|^2/|h_0|^2$. Typical magnitudes are order-of-magnitude and strongly geometry-, frequency-, and loss-dependent.}
\label{tab:taxonomy}
\begin{tabular}{p{0.24\textwidth} p{0.14\textwidth} p{0.24\textwidth} p{0.32\textwidth}}
\toprule
Scenario class & Dominant & Primary diagnostic metric(s) & Practical signature / common pitfall \\
\midrule
Conducting platform / ground plane (Fig.~\ref{fig:fig3}a) &
SELF &
$Z_{\mathrm{in}}$, $R_{\mathrm{in}}$; $\eta_{\mathrm{rad}}$, realized gain &
dB-level TRP/efficiency gains relative to poor return paths; pitfall: attributing improvement only to ``hemispherical pattern'' without checking efficiency and match. \\

Two-ray near-ground propagation (Fig.~\ref{fig:fig3}b) &
TRANSFER &
$|S_{21}|^2$ (or received power) at fixed $S_{11}$; geometry sensitivity &
strong constructive/destructive swings; pitfall: calling it ``gain change'' when Tx/Rx impedances are essentially unchanged. \\

Human body proximity / ``fob-to-head'' (Fig.~\ref{fig:fig3}c) &
MIXED &
$Z_{\mathrm{in}}$ shift + efficiency change (SELF) and link swing (TRANSFER) &
net effect can be positive or negative; pitfall: $R_{\mathrm{in}}\uparrow$ does not imply $R_{\mathrm{rad}}\uparrow$ (absorption may dominate). \\

Handheld chassis/ground excitation (Fig.~\ref{fig:fig3}d) &
SELF &
efficiency/TRP, bandwidth, realized gain; chassis current maps (sim/near-field scans) &
radiation improves by recruiting chassis modes; pitfall: bandwidth increase alone does not guarantee higher $\eta_{\mathrm{rad}}$. \\

Trees / large natural structures (Fig.~\ref{fig:fig4}a) &
SELF (often mixed) &
$Z_{\mathrm{in}}$ and efficiency vs moisture/grounding; received-power variability &
potentially large dB gains in field expedients; pitfall: uncontrolled $R_{\mathrm{loss}}$ and geometry variability limit repeatability. \\

Mountains / reflectors / passive repeaters (Fig.~\ref{fig:fig4}b) &
TRANSFER &
link gain (received power), angular alignment, effective aperture &
enables NLoS links without power; pitfall: attributing to Tx ``pattern improvement'' rather than an added scattering channel. \\

Canyons, corridors, tunnels (Fig.~\ref{fig:fig4}c) &
TRANSFER &
path-loss exponent/intercept, polarization dependence, fading statistics &
confinement can extend coverage but introduces modal fading; pitfall: ignoring mode/polarization selectivity. \\

Engineered surfaces: reflectarrays/metasurfaces/RIS (Fig.~\ref{fig:fig4}d) &
TRANSFER (sometimes mixed) &
calibrated channel gain vs configuration; aperture- and regime-aware scaling &
programmable coupling enhancement/suppression; pitfall: quoting ideal coherent scaling outside its near-/far-field or aperture-limited regime. \\
\bottomrule
\end{tabular}
\end{table}

\subsection{Ground planes and platforms: self Purcell enhancement via radiative damping control}
Figure~\ref{fig:fig3}(a) illustrates the canonical platform mechanism. Conducting platforms (vehicle roofs, bonded ground plates, metallic chassis) are prototypical \emph{local} environments: they modify the self-interaction $\bm{G}(\bm{r}_t,\bm{r}_t)$ and therefore the dissipative overlap integral (Eq.~\eqref{eq:Pavg_ImG}), producing a measurable change in $Z_{\mathrm{in}}$ and, in particular, in the partition $R_{\mathrm{in}}=R_{\mathrm{rad}}+R_{\mathrm{loss}}$ (Eq.~\eqref{eq:port_power_impedance}). In Purcell-first terms, a well-bonded platform increases the radiative part $R_{\mathrm{rad}}$ relative to poor/floating return-path baselines, yielding $\Delta R_{\mathrm{rad}}>0$ and therefore a larger self enhancement under controlled excitation \cite{greffet2010,krasnok2015srep,stanfield2023}. Image-theory language is a geometric way to visualize the same scattering contribution $\bm{G}_{\mathrm{sc}}$, but the causal hierarchy is: strengthened radiative damping first, then redistribution among directions/polarizations/modes \cite{chpinoptics1978,sipe1987,joulain2003,barnes2020jopt}. Two practical caveats matter in the field: finite ground conductivity can simultaneously increase $R_{\mathrm{loss}}$ and reshape propagation (space/surface waves) \cite{norton1937}, and installation quality often sets the baseline (good bonding improves both radiative damping and matching). \cite{balanis2016,kraus2002,stutzman2012}

\subsection{Remote keyless entry (RKE): transfer Purcell factor from environment-shaped coupling}
Figure~\ref{fig:fig3}(b) illustrates a transfer-dominated archetype. In many RKE deployments, the near-ground environment reshapes the end-to-end coupling primarily through the channel coefficient $h$ rather than through a large change in the tiny transmitter’s $Z_{\mathrm{in}}$. Open-area parking lots often approximate a direct-plus-ground-reflected two-ray channel (Section~2), producing strong constructive/destructive swings and a breakpoint beyond which the envelope approaches the $R^{-4}$ regime used for conservative bounds \cite{itu1411,itu525}. A key diagnostic is that large received-power swings can occur with minimal change in $S_{11}$ at the transmitter, i.e., nearly constant $G_{t,\mathrm{real}}$ but strongly varying $F_{\mathrm{tr}}$ \cite{analog2006rke,microchipAN9144}. This is why ``pattern'' intuition can be misleading: the dominant lever is channel coupling.

\subsection{The ``key fob to the head'' effect: mixed self and transfer contributions}
Figure~\ref{fig:fig3}(c) illustrates a mixed case that is naturally interpreted through the factorization in Eq.~\eqref{eq:Pr_ratio_factorization}. The near-ground environment already provides transfer sensitivity (two-ray), while the body introduces a local perturbation that can change $Z_{\mathrm{in}}$ (detuning/matching) and the partition $R_{\mathrm{in}}=R_{\mathrm{rad}}+R_{\mathrm{loss}}$ \cite{hallhao2006,hao2008bodycentric}. Because the body is lossy, $\Delta R_{\mathrm{loss}}$ is typically non-negligible, so $R_{\mathrm{in}}/R_{\mathrm{in},0}$ is not, by itself, a reliable proxy for radiative enhancement unless efficiency is assessed \cite{greffet2010,koenderink2010ol,carminati2015ssr}. Operationally, the body can improve matching (raising delivered power), increase absorption (reducing $\eta_{\mathrm{rad}}$), and perturb the effective two-ray phasing by scattering---so the net outcome depends on how these changes propagate through Eq.~\eqref{eq:Pr_ratio_factorization}.

\subsection{Handheld device ground planes: chassis-mode self enhancement in the small-antenna regime}
Figure~\ref{fig:fig3}(d) highlights the small-antenna regime where recruiting chassis currents is often the dominant route to higher $R_{\mathrm{rad}}$. At sub-GHz frequencies, handheld devices are typically electrically small, so radiative damping is intrinsically weak and constrained by fundamental limits \cite{wheeler1947,chu1948,mclean1996}. In this regime the chassis/ground plane is frequently the dominant current-supporting structure. Designs that more effectively excite chassis modes increase $R_{\mathrm{rad}}$ and $\eta_{\mathrm{rad}}$, and therefore improve realized gain (and often bandwidth), mapping directly to range once the propagation regime is identified \cite{anguera2012groundplane}.

\subsection{Trees as antennas: environment-supplied radiative channels in field expedients}
Figure~\ref{fig:fig4}(a) illustrates the field-expedient ``environmental radiator.'' A tree can supply a current-supporting structure orders of magnitude larger than a portable whip, so the coupled source--environment system can exhibit a much larger effective $R_{\mathrm{rad}}$ and sometimes improved matching \cite{ikrath1975trees}. The defining caveat is variability: moisture content, grounding, and near-field loss mechanisms can strongly change $R_{\mathrm{loss}}$, so received-signal improvements inferred from a single geometry can mix self and transfer contributions (Eq.~\eqref{eq:Pr_ratio_factorization}). This category is therefore best treated as a vivid illustration of how recruiting large current paths can increase radiative damping, while emphasizing the need to assess efficiency when losses are significant.

Tree-coupled “environmental radiators” have been demonstrated most clearly in the MF/HF regime,
where the available current path lengths in the trunk/canopy/roots are electrically significant.
For example, toroid-coupled live trees (HEMAC-style coupling) were reported to outperform a nearby
whip radiator by up to $\sim$20~dB in some cases, with field tests including $\sim$4--5~MHz operation
(and multi-mile links at modest transmitter power) and MF operation around $\sim$425--460~kHz
(with tens-of-miles links reported for large trees) \cite{ikrath1975trees}.

\subsection{Mountains, reflectors, and passive repeaters: transfer enhancement via aperture capture and re-radiation}
Figure~\ref{fig:fig4}(b) illustrates a transfer-dominant class: reflectors and passive repeaters add a strong scattering-mediated channel to the link ($h\approx h_0+h_{\mathrm{sc}}$), increasing $F_{\mathrm{tr}}$ and enabling NLoS connectivity. Mountain reflections can be modeled as an additional coherent contribution under favorable geometry \cite{maeyama1993mountain}, while engineered passive repeaters capture energy with a large effective aperture and re-radiate it toward the receiver \cite{thrower1969radiomirrors,microwavejournal2008passive}. Retrodirective reflectors (e.g., Van Atta arrays) provide robust angular response by returning energy toward the direction of incidence \cite{vanatta1959,appel-hansen1966,larsen1966}. Their performance should be reported as a transfer gain (change in $F_{\mathrm{tr}}$), not as an intrinsic ``pattern'' change of a single antenna.

Mountain-ridge “obstacle gain” and ridge-mediated NLoS coverage are predominantly VHF/UHF phenomena. Classic long-term measurements over a 223~km knife-edge obstacle path used 100 and 751~MHz and showed that transmission loss levels were broadly consistent with knife-edge diffraction theory (after accounting for ridge rounding), with the practical implication that mountain ridges can \emph{enhance} field strength in nominally shadowed regions and reduce loss relative to what would be expected from alternative long-range mechanisms (e.g., scatter) \cite{barsis1961knifeedge}. For engineering predictions, the current ITU-R diffraction recommendation provides the standard knife-edge/rounded-obstacle framework (noting that isolated-obstacle treatments mainly apply to VHF and shorter waves, $f>30$~MHz) \cite{itur_p526_16}.

\subsection{Canyons and tunnels: guided-mode transfer enhancement and mode filtering}
Figure~\ref{fig:fig4}(c) illustrates confinement-driven transfer enhancement. Urban canyons, corridors, and tunnels can support guided or quasi-guided propagation where multiple reflections and waveguide modes confine energy and reshape attenuation with distance relative to open-area propagation \cite{dudley2007tunnels,emslie1975tunnel,zhou2016tunnel}. In link-budget terms, the dominant effect is a modification of $|h|^2$ (and often the effective path-loss exponent over a range), with strong dependence on polarization and modal fading; accordingly, the most diagnostic observables are distance dependence, polarization, and fading statistics rather than the standalone antenna pattern.

Underground corridors and tunnels can behave as oversized, lossy waveguides once the wavelength is smaller than
the transverse dimensions.
Theory and measurements for mine tunnels emphasize UHF operation (e.g., $\sim$200--4000~MHz), where multiple
waveguide modes may propagate and wall roughness/tilt can redistribute energy among modes \cite{emslie1975tunnel}.
Modern tunnel studies spanning 1--10~GHz report that far-field attenuation in tunnels is \emph{generally lower than free-space},
i.e., received power can decay more slowly with distance than the Friis baseline \cite{gerasimov2021tunnels}.
(\emph{Inference for rule-of-thumb magnitude:} if the effective path-loss exponent drops from $n=2$ toward $n\approx 1$--$1.6$,
then over hundreds of meters the deviation from free-space can readily reach $\mathcal{O}(10)$~dB.)

\subsection{Engineered environments: metasurfaces, reflectarrays, and reconfigurable intelligent surfaces (RIS)}
Figure~\ref{fig:fig4}(d) illustrates programmable transfer enhancement. Metasurfaces and RIS platforms aim to design or program $\bm{G}_{\mathrm{sc}}$ so as to strengthen desired scattering-mediated channels, thereby increasing $F_{\mathrm{tr}}$ at target locations. Metasurface theory establishes how engineered surface impedances and phase discontinuities reshape scattering channels \cite{holloway2012,yu2011,sievenpiper1999}, while reflectarrays provide a mature microwave implementation of large-aperture phase engineering \cite{huangencinar2007}. ``Smart radio environments'' translate these ideas into communications language and motivate physics-based models that reconcile Maxwell/Green-function descriptions with link budgets \cite{direnzo2020jsac,basar2019access,elmossallamy2020tccn}. A key practical point is regime dependence: far-field coherent scaling laws are upper bounds that can saturate in near-field or aperture-limited operation, so quantitative claims should be anchored to geometry, wavelength, and the applicable field regime \cite{bjornson2020ojcoms}.


\section{Conclusions and outlook}
This work sharpened a common but often implicit statement: many “environmental enhancements” in antenna systems are best understood as modifications of the \emph{dissipative electromagnetic channels} accessible to a driven source, encoded by the environment-dependent dyadic Green tensor $\bm{G}=\bm{G}_0+\bm{G}_{\mathrm{sc}}$ and, in particular, by how $\Im\{\bm{G}\}$ enters the time-averaged work done by a current distribution.  In the port viewpoint, the same physics appears as an environment-induced change of the real input impedance, $R_{\mathrm{in}}=R_{\mathrm{rad}}+R_{\mathrm{loss}}$, i.e., a change of radiative damping and/or absorption.  This “Purcell-first” hierarchy is fully consistent with classical reaction/reciprocity foundations \cite{purcell1946,rumsey1954,richmond1961} and with LDOS-based nano-optics formulations \cite{sipe1987,joulain2003,carminati2015ssr}.

\paragraph{Positioning relative to existing literature.}
Antenna-based Purcell analogies have already established that, for dipole-like radiators, Purcell factors can be inferred from impedance/radiation-resistance changes and measured in microwave experiments \cite{krasnok2015srep,stanfield2023srep}.  In parallel, the nano-optics LDOS/CDOS literature provides a rigorous and general operator description of both local damping (LDOS) and two-point connectivity (CDOS) \cite{joulain2003,caze2013prl,carminati2015ssr}, and pedagogical accounts have emphasized the deep unity between classical antennas and quantum emitters under the Green-function viewpoint \cite{barnes2020jopt}.  The main contribution of the present manuscript is to make these connections \emph{operational at the link level}: we provide a minimal two-factor factorization of received-power changes into (i) a \emph{self} term tied to measurable port quantities and realized gain and (ii) a \emph{transfer} term tied to channel coupling, together with measurement-aware recipes and falsification tests that prevent the most common misattributions in field practice.

\paragraph{Self vs.\ transfer as the organizing principle.}
A key outcome is the separation of \emph{local (self)} and \emph{link-level (transfer)} Purcell-like effects.  Self effects are governed by the source self-interaction $\bm{G}(\bm{r}_t,\bm{r}_t)$ and are naturally quantified by a radiative-damping ratio such as
$F_{\mathrm{self}}^{(I)}=R_{\mathrm{rad}}/R_{\mathrm{rad},0}$ under a stated excitation convention.  Transfer effects are governed by the cross interaction $\bm{G}(\bm{r}_r,\bm{r}_t)$ (equivalently, mutual coupling/impedance or a normalized channel coefficient) and are naturally quantified by a coupling ratio such as $F_{\mathrm{tr}}=|h|^2/|h_0|^2$ (or, in optics language, by CDOS-type two-point connectivity) \cite{caze2013prl,carminati2015ssr}.  This distinction resolves a pervasive ambiguity: a received-power change can arise from altered realized gains (self: matching, efficiency, directivity) or from altered channel structure (transfer: interference, guiding, scattering), and “pattern improvement” language is often insufficient to diagnose which mechanism dominates.

\paragraph{Implications for interpreting real environments.}
The taxonomy assembled here shows that many widely reported “range tricks” are transfer-dominated (two-ray ground bounce, canyon/tunnel confinement, passive reflectors/repeaters), while many installation and proximity phenomena are self-dominated (platform/ground-plane return paths, chassis-mode excitation, some forms of body loading).  Mixed cases are common, and correct interpretation requires mapping observations back onto a received-power factorization into realized-gain and channel terms, then translating dB-level self or transfer changes into operational range/coverage using regime-appropriate propagation scalings \cite{itu525,itu1411}.

\paragraph{Reproducible reporting: the minimum bar.}
From an experimental and engineering standpoint, the main practical recommendation is to report environmental enhancement using metrics that are robust to loss and source-model ambiguity.  At minimum, a self claim should (i) state the excitation convention, and (ii) separate radiative and dissipative contributions (e.g., by reporting $Z_{\mathrm{in}}$ together with a radiation-efficiency estimate, or by reporting realized gain with explicit mismatch correction).  A transfer claim should normalize out local antenna changes (or monitor $S_{11}$ simultaneously) so that an observed improvement can be attributed to the environment-shaped channel rather than to accidental detuning or matching.  Standardized measurement and definition frameworks provide natural anchors for reproducible reporting \cite{ieee1452013,ieee1492021,ctiaota2022}. See Appendix \ref{app:reproducibility} for more details.

\paragraph{Outlook.}
Three directions appear especially promising.  First, a community-wide \emph{self/transfer scorecard}---reporting $R_{\mathrm{rad}}$, $R_{\mathrm{loss}}$, matching, realized gain, and a calibrated $F_{\mathrm{tr}}$ under a stated geometry and excitation convention---would make results comparable across antennas, propagation, and RIS communities.  Second, engineered environments (reflectarrays, metasurfaces, RIS) offer an ideal testbed for separating and jointly optimizing self and transfer effects, while forcing careful attention to aperture limits and near-/far-field regime dependence \cite{direnzo2020jsac,bjornson2020ojcoms}.  Third, time-varying and programmable environments motivate a dynamic Green-function viewpoint in which $\bm{G}_{\mathrm{sc}}$ is actively modulated, opening routes to robust coupling control, interference management, and adaptive coverage beyond static beam shaping.  By explicitly tying “Purcell language” to measurable impedance and channel quantities, the framework here converts rules of thumb into quantitative, falsifiable principles for practical antenna systems.

\section*{Acknowledgments}
The author acknowledges financial support from the U.S. Department of Energy (DoE) and the U.S. Air Force Office of Scientific Research (AFOSR).

\section*{Competing interests}
The authors declare no competing interests.

\appendix
\section{Reproducibility: extracting $F_{\mathrm{self}}$ and $F_{\mathrm{tr}}$ from measurements and simulations}
\label{app:reproducibility}

\subsection{Definitions, reference planes, and conventions}
This appendix provides a step-by-step recipe to compute the \emph{self} factor $F_{\mathrm{self}}$ and the \emph{transfer} factor
$F_{\mathrm{tr}}$ from either measurements or simulations.
We follow standard antenna definitions for gain, realized gain, and impedance-mismatch factor \cite{ieee1452013},
and standard antenna test procedures where relevant \cite{ieee1491979}.

\paragraph{VNA reference plane.}
All port quantities ($S_{11}$, $Z_{\mathrm{in}}$) must be referenced to the antenna feed terminals via calibration/de-embedding.
Let $Z_0$ denote the VNA reference impedance (typically $50~\Omega$).

\paragraph{From $S_{11}$ to input impedance and mismatch factor.}
Given a one-port measurement $S_{11}(f)=\Gamma(f)$,
\begin{equation}
Z_{\mathrm{in}}(f)=Z_0\,\frac{1+\Gamma(f)}{1-\Gamma(f)} ,
\label{eq:zin_from_s11}
\end{equation}
and the \emph{impedance mismatch factor} (power accepted divided by power incident at the antenna terminals) is
\begin{equation}
\chi(f)\equiv 1-|\Gamma(f)|^{2}.
\label{eq:mismatch_factor}
\end{equation}

\paragraph{Radiation efficiency and realized gain.}
Near resonance, the real part of the input impedance can be written
\begin{equation}
\Re\{Z_{\mathrm{in}}\}=R_{\mathrm{rad}}+R_{\mathrm{loss}},
\label{eq:rin_split}
\end{equation}
where $R_{\mathrm{rad}}$ represents radiative damping and $R_{\mathrm{loss}}$ captures dissipative losses.
The \emph{radiation efficiency} is
\begin{equation}
\eta_{\mathrm{rad}}(f)\equiv \frac{P_{\mathrm{rad}}}{P_{\mathrm{acc}}}
=\frac{R_{\mathrm{rad}}}{R_{\mathrm{rad}}+R_{\mathrm{loss}}}.
\label{eq:eta_rad_def}
\end{equation}
With directivity $D(\hat{\mathbf{r}},f)$, the (absolute) gain is $G=D\,\eta_{\mathrm{rad}}$ and the \emph{realized gain} is
\begin{equation}
G_{\mathrm{real}}(\hat{\mathbf{r}},f)=G(\hat{\mathbf{r}},f)\,\chi(f)
= D(\hat{\mathbf{r}},f)\,\eta_{\mathrm{rad}}(f)\,\bigl(1-|\Gamma(f)|^2\bigr).
\label{eq:realized_gain}
\end{equation}
This convention is consistent with IEEE definitions: gain uses \emph{accepted} power, while realized gain is reduced by mismatch \cite{ieee1452013}.

\paragraph{Self and transfer factors (operational definitions).}
In the constant-current convention emphasized in the main text, the self factor is
\begin{equation}
F_{\mathrm{self}}^{(I)}(f)\equiv \frac{R_{\mathrm{rad}}^{(\mathrm{env})}(f)}{R_{\mathrm{rad}}^{(0)}(f)}
=\frac{P_{\mathrm{rad}}^{(\mathrm{env})}}{P_{\mathrm{rad}}^{(0)}} \quad (\text{same feed current}).
\label{eq:Fself_I_def}
\end{equation}
The transfer factor is defined by normalizing the measured link to a chosen propagation baseline:
\begin{equation}
F_{\mathrm{tr}}(f)\equiv \frac{|h^{(\mathrm{env})}(f)|^2}{|h^{(0)}(f)|^2}
\;\;\Longleftrightarrow\;\;
\frac{P_r}{P_t} = \Bigl[G_{\mathrm{real},t}(\hat{\mathbf{r}}_t)G_{\mathrm{real},r}(\hat{\mathbf{r}}_r)\Bigr]\,
\Bigl(\frac{\lambda}{4\pi d}\Bigr)^2\,F_{\mathrm{tr}},
\label{eq:Ftr_link_def}
\end{equation}
where the baseline $(\lambda/4\pi d)^2$ corresponds to free-space propagation (Friis) \cite{friis1946}
and can be replaced by an ITU short-range model when appropriate (Sec.~\ref{app:baseline_models}).

\subsection{Recipe A: from VNA measurements ($S_{11}\rightarrow Z_{\mathrm{in}}$, $\eta_{\mathrm{rad}}$, $G_{\mathrm{real}}$)}
\label{app:recipe_vna}

\begin{enumerate}
\item \textbf{Calibrate to the feed.}
Perform SOLT/TRL calibration (and fixture/cable de-embedding) so that $S_{11}$ is referenced to the antenna terminals.

\item \textbf{Measure $S_{11}(f)$ and compute $Z_{\mathrm{in}}(f)$.}
Use Eqs.~\eqref{eq:zin_from_s11}--\eqref{eq:mismatch_factor} to obtain $Z_{\mathrm{in}}(f)$ and $\chi(f)$.

\item \textbf{Choose the frequency point(s).}
For narrowband antennas, report values at (i) the resonance $f_0$ (where $\Im\{Z_{\mathrm{in}}\}\approx 0$)
and (ii) any additional frequencies used for link tests.

\item \textbf{Estimate radiation efficiency $\eta_{\mathrm{rad}}(f)$ (choose one method).}
\begin{enumerate}
\item \emph{Wheeler-cap (differential impedance) method.}
Measure $S_{11}^{(\mathrm{fs})}(f)$ with radiation allowed and $S_{11}^{(\mathrm{cap})}(f)$ with the antenna inside a conductive cap
that suppresses radiation.
Convert both to impedances and evaluate (near resonance)
\begin{equation}
R_{\mathrm{in}}^{(\mathrm{fs})}=\Re\{Z_{\mathrm{in}}^{(\mathrm{fs})}\},\qquad
R_{\mathrm{loss}}\approx \Re\{Z_{\mathrm{in}}^{(\mathrm{cap})}\},
\end{equation}
so that
\begin{equation}
R_{\mathrm{rad}}\approx R_{\mathrm{in}}^{(\mathrm{fs})}-R_{\mathrm{loss}},
\qquad
\eta_{\mathrm{rad}}\approx \frac{R_{\mathrm{rad}}}{R_{\mathrm{in}}^{(\mathrm{fs})}}
=1-\frac{R_{\mathrm{loss}}}{R_{\mathrm{in}}^{(\mathrm{fs})}}.
\label{eq:wheeler_eta}
\end{equation}
This method is most reliable for electrically small / single-mode antennas and requires care to avoid cap loading artifacts \cite{stefanelli2011ursi,moharram2014ursi}.

\item \emph{Reverberation-chamber (RC) methods.}
Measure \emph{total efficiency} $\eta_{\mathrm{tot}} \equiv P_{\mathrm{rad}}/P_{\mathrm{av}}$ and/or \emph{radiation efficiency}
$\eta_{\mathrm{rad}} \equiv P_{\mathrm{rad}}/P_{\mathrm{acc}}$ following established RC procedures \cite{holloway2012rc}.
When only $\eta_{\mathrm{tot}}$ is obtained, correct mismatch using
\begin{equation}
\eta_{\mathrm{tot}} = \eta_{\mathrm{rad}}\,\chi
\quad\Rightarrow\quad
\eta_{\mathrm{rad}} = \eta_{\mathrm{tot}}/\chi,
\label{eq:eta_tot_relation}
\end{equation}
consistent with the distinction between available and accepted power at the port \cite{holloway2012rc,ieee1452013}.

\item \emph{Pattern integration / gain-transfer method.}
Measure the gain (or realized gain) pattern in an anechoic range and compute directivity by pattern integration.
Then use $\eta_{\mathrm{rad}} = G/D$ and $G_{\mathrm{real}}=G\,\chi$ (or directly report $G_{\mathrm{real}}$ if measured as such),
following standard antenna test practice \cite{ieee1491979}.
\end{enumerate}

\item \textbf{Compute realized gain.}
Once $D(\hat{\mathbf{r}},f)$ and $\eta_{\mathrm{rad}}(f)$ are known/estimated,
compute $G_{\mathrm{real}}$ from Eq.~\eqref{eq:realized_gain}.
Report explicitly whether $D$ is simulated (common) while $\eta_{\mathrm{rad}}$ is measured (common for lossy antennas).

\end{enumerate}

\subsection{Recipe B: extracting $F_{\mathrm{self}}$ from measurements/simulations}
\label{app:recipe_fself}

\paragraph{Measurement-based $F_{\mathrm{self}}^{(I)}$.}
Repeat the VNA/efficiency procedure in two configurations:
(i) a reference condition ``0'' (e.g., free space / anechoic), and
(ii) the environment of interest (``env'').
At (or near) resonance, extract $R_{\mathrm{rad}}$ using either the Wheeler-cap split (Eq.~\eqref{eq:wheeler_eta})
or the identity implied by Eqs.~\eqref{eq:rin_split}--\eqref{eq:eta_rad_def}:
\begin{equation}
R_{\mathrm{rad}}(f)\approx \eta_{\mathrm{rad}}(f)\,\Re\{Z_{\mathrm{in}}(f)\}.
\label{eq:Rrad_from_eta}
\end{equation}
Then compute
\begin{equation}
F_{\mathrm{self}}^{(I)}(f)=\frac{R_{\mathrm{rad}}^{(\mathrm{env})}(f)}{R_{\mathrm{rad}}^{(0)}(f)}.
\end{equation}
A key falsification check is: if received power changes by many dB while $\Re\{Z_{\mathrm{in}}\}$, $|\Gamma|$, and delivered power
remain unchanged (within uncertainty), then the observation is transfer-dominant rather than a self enhancement.

\paragraph{Simulation-based $F_{\mathrm{self}}^{(I)}$.}
In a full-wave solver (MoM/FEM/FDTD), for each configuration compute:
accepted port power $P_{\mathrm{acc}}$, radiated power $P_{\mathrm{rad}}$ (far-field flux integration),
and $\Gamma=S_{11}$ at the feed.
Then $\eta_{\mathrm{rad}}=P_{\mathrm{rad}}/P_{\mathrm{acc}}$ and $R_{\mathrm{rad}}$ can be obtained either by
(i) extracting the feed current $I$ and using $R_{\mathrm{rad}}=2P_{\mathrm{rad}}/|I|^2$ (constant-current convention),
or (ii) using Eq.~\eqref{eq:Rrad_from_eta} at resonance.
Finally compute the ratio in Eq.~\eqref{eq:Fself_I_def}.

\subsection{Recipe C: extracting $F_{\mathrm{tr}}$ from a link test}
\label{app:recipe_ftr}

Assume a narrowband tone (or sufficiently narrow subband) at frequency $f$ and separation $d$.
Measure:
(i) transmit \emph{available} power (or forward power at the calibrated plane),
(ii) $S_{11,t}$ and $S_{11,r}$ (or $\Gamma_t,\Gamma_r$) for mismatch,
(iii) received power $P_r$ at the receiver (with cable losses calibrated out), and
(iv) realized gains $G_{\mathrm{real},t}$ and $G_{\mathrm{real},r}$ in the link directions.

\paragraph{Step-by-step.}
\begin{enumerate}
\item \textbf{Fix the power reference.}
If you measure forward power $P_{\mathrm{fwd}}$ at the transmit port, compute accepted power
$P_{t,\mathrm{acc}} = P_{\mathrm{fwd}}\,\chi_t$ with $\chi_t=1-|S_{11,t}|^2$.

\item \textbf{Use realized gains.}
Compute or measure $G_{\mathrm{real},t}$ and $G_{\mathrm{real},r}$ at the link pointing angles
(Eq.~\eqref{eq:realized_gain}). If only $G$ is available, convert via $G_{\mathrm{real}}=G\,\chi$.

\item \textbf{Compute the free-space reference received power.}
Using Friis (free-space baseline),
\begin{equation}
P_{r}^{(0)} = P_{t,\mathrm{acc}}\,G_{t}(\hat{\mathbf{r}}_t)\,G_{r}(\hat{\mathbf{r}}_r)\,\Bigl(\frac{\lambda}{4\pi d}\Bigr)^2,
\end{equation}
or equivalently
\begin{equation}
P_{r}^{(0)} = P_{t,\mathrm{av}}\,G_{\mathrm{real},t}(\hat{\mathbf{r}}_t)\,G_{\mathrm{real},r}(\hat{\mathbf{r}}_r)\,\Bigl(\frac{\lambda}{4\pi d}\Bigr)^2,
\label{eq:Pr0_realized}
\end{equation}
depending on whether you reference transmit power to accepted or available power \cite{friis1946,ieee1452013}.

\item \textbf{Extract the transfer factor.}
\begin{equation}
F_{\mathrm{tr}} = \frac{P_r}{P_{r}^{(0)}}.
\label{eq:Ftr_from_link}
\end{equation}
In dB form (using realized gains and ITU free-space loss $L_{\mathrm{bf}}$),
\begin{equation}
10\log_{10}F_{\mathrm{tr}}
= \bigl[P_r - P_{t,\mathrm{av}} - G_{\mathrm{real},t} - G_{\mathrm{real},r}\bigr]_{\mathrm{dB}}
+ L_{\mathrm{bf}}(f,d),
\label{eq:Ftr_dB}
\end{equation}
where $L_{\mathrm{bf}}$ is the free-space basic transmission loss (Sec.~\ref{app:baseline_models}).
\end{enumerate}

\paragraph{Practical notes.}
(i) Polarization mismatch is not included in realized gain and should be controlled or reported.
(ii) Ensure the baseline is applied in its validity regime (far field, unobstructed LoS) or replace the baseline with a short-range ITU model as below.
(iii) If $S_{11}$ is unchanged while $P_r$ varies strongly with position/time, the change should be attributed primarily to transfer (multipath/fading) rather than self.

\subsection{Choosing the propagation baseline: ITU free-space vs.\ short-range models}
\label{app:baseline_models}

\paragraph{Free-space baseline (ITU-R P.525).}
Use free-space loss as the baseline when the link is approximately unobstructed line-of-sight and the antennas operate in (or near) each other’s far field.
ITU-R P.525 defines the free-space basic transmission loss between isotropic antennas as
\begin{equation}
L_{\mathrm{bf}}(f,d)=20\log_{10}\Bigl(\frac{4\pi d}{\lambda}\Bigr)
=32.4+20\log_{10}f_{\mathrm{MHz}}+20\log_{10}d_{\mathrm{km}}\;\;\mathrm{dB}.
\label{eq:itu_p525}
\end{equation}
This is the natural baseline for $F_{\mathrm{tr}}$ when the ``null hypothesis'' is free-space propagation \cite{itur_p525_4}.

\paragraph{Short-range outdoor baseline (ITU-R P.1411).}
For outdoor links in built environments where buildings/streets dominate propagation and multipath/fading are intrinsic (typical distances $\lesssim 1$~km),
use ITU-R P.1411, which provides basic transmission loss models for LoS and NLoS and associated fading/multipath characteristics over 300~MHz--100~GHz \cite{itur_p1411_13}.
In this case, define $F_{\mathrm{tr}}$ relative to the P.1411-predicted median loss for the appropriate scenario rather than Eq.~\eqref{eq:itu_p525}.

\bibliographystyle{unsrt}
\bibliography{references}
\end{document}